\pdfoutput=1
\RequirePackage{ifpdf}
\ifpdf 
\documentclass[pdftex]{sigma}
\else
\documentclass{sigma}
\fi

\numberwithin{equation}{section}

\newtheorem{Theorem}{Theorem}[section]

\newtheorem{Proposition}[Theorem]{Proposition}
 { \theoremstyle{definition}

 }

\usepackage{mathtools}
\usepackage{makecell}
\usepackage{tikz}
\usepackage{tikz-cd}
\usetikzlibrary{arrows}
\usetikzlibrary{decorations.markings}
\tikzset{->-/.style={decoration={
 markings,
 mark=at position #1 with {\arrow{>}}}, postaction={decorate}}}
\tikzset{->>-/.style={decoration={
 markings,
 mark=at position #1 with {\arrow{>>}}}, postaction={decorate}}}

\usepackage[cal=boondox,bb=boondox]{mathalfa}

\DeclareMathOperator{\Tr}{Tr}

\begin{document}
\allowdisplaybreaks

\newcommand{\arXivNumber}{1711.03379}

\renewcommand{\PaperNumber}{003}

\FirstPageHeading

\ShortArticleName{Note on Character Varieties and Cluster Algebras}

\ArticleName{Note on Character Varieties and Cluster Algebras}

\Author{Kazuhiro HIKAMI}

\AuthorNameForHeading{K.~Hikami}

\Address{Faculty of Mathematics, Kyushu University, Fukuoka 819-0395, Japan}
\Email{\href{mailto:khikami@gmail.com}{khikami@gmail.com}}

\ArticleDates{Received July 25, 2018, in final form January 10, 2019; Published online January 20, 2019}

\Abstract{We use Bonahon--Wong's trace map to study character varieties of the once-punctured torus and of the $4$-punctured sphere. We clarify a relationship with cluster algebra associated with ideal triangulations of surfaces, and we show that the Goldman Poisson algebra of loops on surfaces is recovered from the Poisson structure of cluster algebra. It is also shown that cluster mutations give the automorphism of the character varieties. Motivated by a work of Chekhov--Mazzocco--Rubtsov, we revisit confluences of punctures on sphere from cluster algebraic viewpoint, and we obtain associated affine cubic surfaces constructed by van~der~Put--Saito based on the Riemann--Hilbert correspondence. Further studied are quantizations of character varieties by use of quantum cluster algebra.}


\Classification{13F60; 30F60; 33E17; 57Q15}

\section{Introduction}
The Kauffman bracket skein algebra was used to introduce a quantization of $\mathrm{SL}(2;\mathbb{C})$ character variety of surface~\cite{PrzytSikor00a}. The skein algebra was also regarded as a quantization of the Goldman Poisson algebras of loops on surfaces~\cite{VGTura91b}. In~\cite{BonaWong11a,BonaWong16b}~(see also~\cite{AllegHKKim17a, TQLe16a}) Bonahon and Wong introduced a quantum trace map of
$\mathrm{SL}(2;\mathbb{C})$ representations of surface groups, and defined a map from the skein algebra to the quantum Teichm{\"u}ller space introduced in~\cite{ChekFock99a, Kasha98a} based on ideal triangulations of surfaces.

The cluster algebra was introduced by Fomin and Zelevinsky~\cite{FominZelev02a}, and it is known that the quantum Teichm{\"u}ller space has a cluster algebraic structure~\cite{FockGonc06b,GekhShapVain05a}. Namely
the Ptolemy type relation for the $\lambda$-length (see~\cite{Penner12}) is interpreted as a cluster mutation, and a corresponding quantized cluster mutation assigns a quantum dilogarithm function to a flip of triangulations studied in~\cite{Kasha98a}. A formulation of triangulated surface in terms of the cluster algebra was further investigated in~\cite{FomiShapThur08a}, and it gives a class of finite-mutation-type cluster algebra.

Purpose of this paper is to apply Bonahon--Wong's trace map to the character varieties of the once-punctured torus and of the $4$-punctured sphere. We employ the cluster algebra, which originates from ideal triangulation of surfaces and was used~\cite{HikamiRInoue12a} for hyperbolic volumes of the once-punctured torus bundle over $S^1$ and the $2$-bridge link complements. We give a~cluster algebra realization of the character varieties (Theorems~\ref{thm:torus_phi} and~\ref{thm:map_sphere}). We also construct the $\mathrm{PGL}(2;\mathbb{Z})$ automorphism group actions on these character varieties by use of the cluster mutations.

It is known that the character variety of the $4$-punctured sphere is the affine cubic surface related to the monodromy preserving transformation of the Painlev{\'e}~VI equation. Other Painlev{\'e} equations have irregular singularities, and the associated affine cubic surfaces analogous to the Fricke surface were given in~\cite{vdPutSaito09a} as the moduli space of the generalized monodromy in the Riemann--Hilbert correspondence.
In~\cite{ChekMazzRubt15a}, those affine surfaces were studied by use of the $\lambda$-length of the Teichm{\"u}ller space, and discussed was a relationship with a \emph{generalized cluster algebra}~\cite{ChekhShapi14a}.
In this article, we revisit degeneration processes of the character variety of the $4$-punctured sphere, and reformulate them using the \emph{cluster algebra}. We propose $11$ families of the finite-mutation-type cluster algebra whose quiver is for a degeneration of ideal triangulation of the punctured sphere (Theorem~\ref{thm:degenerate}). We see that they are intimately related to confluences of the Painlev{\'e} equations, and that the degenerated character varieties agree with those in~\cite{vdPutSaito09a}. We also quantize the affine cubic surface using the quantum cluster algebra~\cite{BerenZelev05a,FockGonc09a}, and we obtain the same results with~\cite{BulloPrzyt99a} which was derived using the skein algebra on the punctured torus (Theorem~\ref{thm:q_torus}) and on the punctured sphere (Theorem~\ref{thm:qW_for_sphere}). Theorem~\ref{thm:q_sphere} is for a quantization of degenerated character varieties. We note that the quantum algebra related to the character varieties of the 4-punctured sphere has a relationship~\cite{Oblom04a} with the $C^\vee C_1$ double affine Hecke algebra of Cherednik~\cite{Chered05Book}, whose polynomial representation gives the Askey--Wilson polynomial (see also~\cite{Koorn08a,Terwil13a}). As the $\mathrm{PSL}(2;\mathbb{Z})$ automorphism of the $C^\vee C_1$ DAHA was used to construct knot invariant related with the categorification~\cite{IChered16a} (see also~\cite{BerestSamuel16a}), our cluster algebraic formulation may help for such invariants~\cite{Hikami}. It should be noted that
the quantum trace map of closed paths could be regarded as the expectation value of a supersymmetric line defect in the $\mathcal{N}=2$ four dimensional theory, which is conjectured to be written in terms of the quantum
Fock--Goncharov coordinates, i.e., the quantum cluster $Y$-variables~\cite{GaioMoorNeit13a} (see also~\cite{MGabel17a}).

This paper is organized as follows. In Section~\ref{sec:cluster} we briefly give definitions of the cluster algebra and its quantization. In Section~\ref{sec:torus} studied is the character variety of the once-punctured torus. We give an explicit form of Bonahon--Wong's trace map in terms of the cluster variables. Also shown is that the automorphism group is written by use of the cluster mutations. In Section~\ref{sec:sphere}, we clarify a relationship between the character variety of the $4$-punctured sphere and the cluster algebra associated with ideal triangulation of the surface. In Section~\ref{sec:confluence} we study degenerations of the character varieties of the $4$-punctured sphere from a viewpoint of cluster algebra. We obtain finite-mutation-type quivers for each degenerated affine cubic surfaces. The automorphisms of surfaces are also given by use of the cluster mutations. Sections~\ref{sec:q_torus}--\ref{sec:q_confluence} are devoted to quantization of the character varieties. We use the quantum cluster algebra to give explicit forms of the quantized affine cubic surfaces.

\section{Cluster algebra} \label{sec:cluster}
\subsection[Quiver, $y$-variables, and mutations]{Quiver, $\boldsymbol{y}$-variables, and mutations}
A quiver $Q$ is an oriented graph with $|Q|$ vertices. We have an exchange matrix $\mathbf{B} = (b_{ij})$ defined by
\begin{gather*}
 b_{ij}= \#\{ \text{arrows from $i$ to $j$}\} - \# \{ \text{arrows from $j$ to $i$} \} ,
\end{gather*}
where~$i$ and~$j$ denote vertices of~$Q$. Assuming that $Q$ has neither loops nor 2-cycles, $\mathbf{B}$ is a~skew-symmetric $|Q|\times |Q|$ integral matrix. Each vertex of $Q$ is assigned a cluster $y$-variable~$y_i$. A~seed is a~pair $(\boldsymbol{y}, \mathbf{B})$, where we mean $\boldsymbol{y}=(y_1, \dots, y_{|Q|})$~\cite{FominZelev02a}. The $y$-variable corresponds to the coefficient in~\cite{FominZelev02a}, and it is related to the $x$-variable.

A mutation $\mu_k$ at the vertex $k$ is a map between two seeds, $(\boldsymbol{y}, \mathbf{B})$ and $(\widetilde{\boldsymbol{y}}, \widetilde{\mathbf{B}})$, defined by
\begin{gather} \label{mutation_y}
 \mu_k( \boldsymbol{y}, \mathbf{B}) = \big(\widetilde{\boldsymbol{y}} , \widetilde{\mathbf{B}}\big) ,
\end{gather}
where
\begin{gather*}
 \widetilde{y}_i
 =
 \begin{cases}
 {y_k}^{-1} , & \text{for $i=k$}, \\
 y_i \big( 1+{y_k}^{-1} \big)^{-b_{ki}} , & \text{for $i \neq k$, $b_{ki} \geq 0$}, \\
 y_i ( 1+y_k )^{-b_{ki}} , & \text{for $i \neq k$, $b_{ki} \leq 0$},
 \end{cases}
\end{gather*}
and
an exchange matrix $ \widetilde{\mathbf{B}}= \big( \widetilde{b}_{ij} \big)$ is
\begin{gather} \label{mutation_B}
 \widetilde{b}_{ij} =
 \begin{cases}
 - b_{ij} , & \text{for $i=k$ or $j=k$}, \\
 b_{ij}+ \frac{1}{2} ( | b_{ik} | b_{kj} + b_{ik} | b_{kj} | ) , & \text{otherwise}.
 \end{cases}
\end{gather}
Note that the mutation is involutive, $\mu_k \circ \mu_k={\rm id}$. Two seeds are called mutation equivalent when they are related by a sequence of mutations. We denote $\mathcal{A}(Q)$ as algebra generated by all cluster
$y$-variables of all mutation-equivalent to the initial seed $(\boldsymbol{y}, \mathbf{B})$, where $\mathbf{B}$ is assigned to a quiver~$Q$. When a set of mutation equivalent seeds is finite, $\mathcal{A}(Q)$ is said to be
finite-type. It was proved~\cite{FominZelev03a} that $\mathcal{A}(Q)$ is finite-type if and only if the quiver has a Dynkin diagram. A~cluster algebra with only finitely many exchange matrices is called finite-mutation-type~\cite{FomiShapThur08a}, and it was classified in~\cite{FeliShapTuma12c}. A quiver, which corresponds to an ideal (or, tagged) triangulation of a~surface~\cite{FomiShapThur08a}, belongs to this type, and
a~cluster mutation is regarded as a flip.

For any seed $(\boldsymbol{y}, \mathbf{B})$, we have a Poisson structure in a log-canonical form (see, e.g.,~\cite{GekhShapVain10Book})
\begin{gather} \label{def_Poisson}
\{ y_j , y_k \} = b_{jk} y_j y_k .
\end{gather}
The cluster mutations~\eqref{mutation_y} preserve the Poisson structure, $\big\{ \widetilde{y}_j , \widetilde{y}_k\big\} =\widetilde{b}_{jk} \widetilde{y}_j \widetilde{y}_k$. We should remark that an inversion of $y$-variable
\begin{gather} \label{inversion_cluster}
 \varsigma(\boldsymbol{y}, \mathbf{B}) = \big(\big ({y_1}^{-1}, \dots, {y_{|Q|}}^{-1}\big), \mathbf{B} \big)
\end{gather}
also preserves the Poisson structure~\eqref{def_Poisson}. This type of the inversion action was also used in~\cite{BersGavrMars18a}.

\subsection{Quantum cluster algebra}
As a quantization of the Poisson algebra~\eqref{def_Poisson}, we introduce a $q$-commuting variable $\boldsymbol{Y}=(Y_1, \dots, Y_{|Q|})$~\cite{FockGonc09b, FockGonc09a} satisfying
\begin{gather} \label{quantum_Y}
 Y_k Y_j = q^{2 b_{jk}} Y_j Y_k .
\end{gather}
A mutation at $k$ is quantized as
\begin{gather*}
 \mu^q_k ( \boldsymbol{Y}, \mathbf{B}) = \big( \widetilde{\boldsymbol{Y}}, \widetilde{ \mathbf{B} }\big) ,
\end{gather*}
where
an exchange matrix~$\widetilde{\mathbf{B}}$ is same to the classical case~\eqref{mutation_B}, and the quantized variables are
\begin{gather*}
 \widetilde{Y}_i= \begin{cases}
 {Y_k}^{-1} , & \text{for $i=k$}, \\
 \displaystyle Y_i \prod_{m=1}^{b_{ki}} \big( 1+ q^{2m-1} {Y_k}^{-1} \big)^{-1} , & \text{for $i\neq k$, $b_{ki} \geq 0$}, \\
 \displaystyle Y_i \prod_{m=1}^{-b_{ki}} \big( 1 + q^{2m-1} Y_k \big) , & \text{for $i \neq k$, $b_{ki} \leq 0$}.
 \end{cases}
\end{gather*}
It is remarked that the quantized mutations preserve the $q$-commutation relations~\eqref{quantum_Y}, $\widetilde{Y}_k \widetilde{Y}_j=q^{2 \widetilde{b}_{jk}} \widetilde{Y}_j \widetilde{Y}_k$. We denote $\mathcal{A}^q(Q)$ as a quantization of the cluster algebra $\mathcal{A}(Q)$ generated by all cluster $Y$-variables which are mutation-equivalent to the initial seed~$(\boldsymbol{Y}, \mathbf{B})$.

\section{Once-punctured torus}\label{sec:torus}
\subsection{Character variety and cubic surface}
\begin{figure}[tbhp] \centering
 \begin{minipage}{0.4\linewidth}
 \centering
 \includegraphics[scale=1]{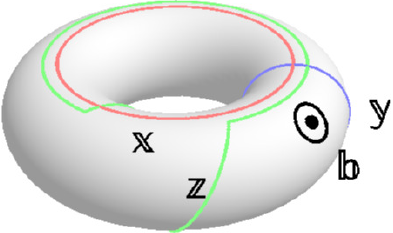}
 \end{minipage}
 \qquad
 \begin{minipage}{0.4\linewidth} \centering
 \includegraphics[scale=1]{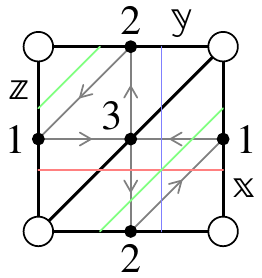}
 \end{minipage}
 \caption{Closed paths on the once-punctured torus $\Sigma_{1,1}$ are depicted in the left figure. The path~$\mathbb{b}$ encircles a puncture. In the right, given is an ideal triangulation of $\Sigma_{1,1}$, whose vertices denote the puncture. Therein we assign a vertex of a quiver $Q_{\text{Mar}}$ on each edge of triangles.} \label{fig:torus}
\end{figure}

Let $\Sigma_{1,1}$ be once-punctured torus. The fundamental group $\pi_1(\Sigma_{1,1})$ is freely generated by simple closed curves $\mathbb{x}$ and $\mathbb{y}$ in Fig.~\ref{fig:torus}. We also define $\mathbb{z}$ and $\mathbb{b}$ as in Fig.~\ref{fig:torus}, and correspondingly we set the $\mathrm{SL}(2;\mathbb{C})$-characters by
\begin{gather*}
 b=\Tr \rho(\mathbb{b}), \qquad x=\Tr \rho(\mathbb{x}), \qquad y=\Tr \rho(\mathbb{y}), \qquad z=\Tr \rho(\mathbb{z}).
\end{gather*}
It is known since Fricke that the moduli space of the equivalence classes of the $\mathrm{SL}(2;\mathbb{C})$ representations is identified with $(x,y,z)\in \mathbb{C}^3$ such that
\begin{gather} \label{W_torus}
 W_{\text{Mar}}(x,y,z;b) = x y z -\big(x^2+y^2+z^2\big)+2 -b = 0.
\end{gather}
We note that the monodromy around the puncture depends on the parameter~$b$. It is a modi\-fication of the Markoff equation, and it receives wide interests from both number theory and hyperbolic geometry (see, e.g.,~\cite{Aigner13Book}).

We define the character variety
\begin{gather} \label{X_torus}
 \mathcal{X}(\Sigma_{1,1}) = \big\{ (x,y,z) \in \mathbb{C}^3 \,|\, W_{\text{Mar}}(x,y,z;b)=0 \big\} ,
\end{gather}
for fixed $b$.

\subsection{Cluster algebra and character variety}
An ideal triangulation of $\Sigma_{1,1}$ is depicted in Fig.~\ref{fig:torus}. Therein we assign vertices of quiver to each edge of the ideal triangles, and give an anti-clockwise orientation to each dual face. Each label is for a vertex of the quiver, and gray lines constitute the following quiver $Q_{\text{Mar}}$ which is known as the Markoff quiver
\begin{gather*}
 Q_{\text{Mar}} =
 \raisebox{-12mm}{
 \begin{tikzpicture}
 \draw [->-=.7] (2,0)to[bend right=15](0,0);
 \draw [->-=.7] (2,0)to[bend left=15](0,0);
 \draw [->-=.7] (0,0)to[bend right=15](1,{sqrt(3)});
 \draw [->-=.7] (0,0)to[bend left=15](1,{sqrt(3)});
 \draw [->-=.7] (1,{sqrt(3)})to[bend right=15](2,0);
 \draw [->-=.7] (1,{sqrt(3)})to[bend left=15](2,0);
 \filldraw (0,0) circle[radius=1.2pt];
 \filldraw (2,0) circle[radius=1.2pt];
 \filldraw (1,{sqrt(3)}) circle[radius=1.2pt];
 \node [above ] at (1,{sqrt(3)}) {$1$};
 \node [below left ] at (0,0) {$2$};
 \node [below right ] at (2,0) {$3$};
 \end{tikzpicture}
 }
\end{gather*}
The exchange matrix is then defined as
\begin{gather} \label{B_torus}
 \mathbf{B}_{\text{Mar}}= \begin{pmatrix}
 0 & -2 & 2
 \\
 2 & 0 & -2
 \\
 -2 & 2 & 0
 \end{pmatrix} .
\end{gather}

The character variety~\eqref{X_torus} is realized from the cluster algebra as follows.
\begin{Theorem} \label{thm:torus_phi} We have an embedding
 \begin{gather*}
 \iota\colon \ \mathcal{X}(\Sigma_{1,1}) \to \mathcal{A}(Q_{\text{\rm Mar}})
 \end{gather*}
 defined by
 \begin{gather}
 \iota(b) = {y_1 y_2 y_3} + \frac{1}{ {y_1 y_2 y_3} } , \nonumber \\
 \iota(x) = \sqrt{y_1 y_3} + \sqrt{\frac{y_1}{y_3}} + \frac{1}{\sqrt{y_1 y_3}} , \qquad \iota(y) = \sqrt{y_2 y_3} + \sqrt{\frac{y_3}{y_2}} + \frac{1}{\sqrt{y_2 y_3}} \nonumber \\
 \iota(z) = \sqrt{y_1 y_2} + \sqrt{\frac{y_2}{y_1}}
 + \frac{1}{\sqrt{y_1 y_2}} . \label{xyz_y}
 \end{gather}
\end{Theorem}
The equality~\eqref{W_torus} can be checked directly from~\eqref{xyz_y}.

\begin{figure} \centering
 \includegraphics[scale=1.]{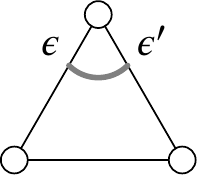}
 \caption{Arc on ideal triangulation, whose Boltzmann weight is $\gamma_\Delta(\epsilon, \epsilon^\prime)$.} \label{fig:arc}
\end{figure}

The definitions~\eqref{xyz_y} were given in~\cite{FockGonc06b,FockGonc07b}. Here we follow the Bonahon--Wong trace map of closed path introduced in~\cite{BonaWong11a} (see also~\cite{TQLe16a}). Let $\Delta$ be one of faces of triangulations of surface~$\Sigma$. For each $\Delta$, we assign triangle algebra $\mathcal{T}_\Delta$ generated by $z_{ 1}$, $z_{ 2}$, and $z_{ 3}$, satisfying the Poisson relation
\begin{gather} \label{triangle_algebra}
\{z_{ j} , z_{ j+1}\}= z_{ j} z_{ j+1} ,
\end{gather}
where $j \in \{1,2,3\}$ is read cyclically. These $z_j$ are assigned to each edge of $\Delta$ in clockwise order. When the triangulation of surface is $\Sigma =\bigcup_j \Delta_j$, we have a tensor product algebra $\bigotimes_j \mathcal{T}_{\Delta_j}$. In addition to $z_j$ we introduce a state $\epsilon\in \{ +1, -1 \}$ to each edge of~$\Delta_j$. In gluing faces $\Delta_j$ together along edge, a state $\epsilon$ on the edge should be compatible. Then the state sum for closed path on~$\Sigma$ is given as a trace of product of Boltzmann weight $\gamma_\Delta(\epsilon, \epsilon^\prime)$, which is assigned to arc in~$\Delta$ homotopic to the arc in Fig.~\ref{fig:arc}
\begin{gather} \label{gamma_arc}
 \gamma_\Delta(\epsilon, \epsilon^\prime) =
 \begin{cases}
 0 , & \text{if $(\epsilon, \epsilon^\prime)=(-1, +1)$}, \\
 {z_{ j}}^{\frac{1}{2}\epsilon} {z_{ j+1}}^{\frac{1}{2}\epsilon^\prime}, & \text{otherwise}.
 \end{cases}
\end{gather}
Here $z_j$ (resp. $z_{j+1}$) is a generator of $\mathcal{T}_\Delta$ assigned to edge of state $\epsilon$ (resp. $\epsilon^\prime$).

In our case of the once-punctured torus, $\Sigma_{1,1}$ is triangulated into $\Delta_1 \cup \Delta_2$, where $\Delta_1$ (resp.~$\Delta_2$) is an upper (resp.\ lower) triangle in Fig.~\ref{fig:torus}. A path $\mathbb{x}$, for instance, is presented as a union of arcs in Fig.~\ref{fig:arc} as
\begin{gather*}
 \sum_{(\epsilon_1, \epsilon_3) \neq (+1, -1)} \gamma_{\Delta_1}(\epsilon_3, \epsilon_1) \otimes \gamma_{\Delta_2}(\epsilon_3, \epsilon_1) .
\end{gather*}
By setting $y_1={z_1}^{-1} \otimes {z_1}^{-1}$ and $y_3={z_3}^{-1} \otimes {z_3}^{-1}$, we get an expression for $\mathbb{x}$ as in~\eqref{xyz_y}. Other paths can be read similarly.

Note that~\eqref{xyz_y} was used in~\cite{FockGonc07b} for a study of the Markoff equation. It should be remarked that $\iota(x)$ suggests that we have up to conjugation
\begin{gather*}
 \rho(\mathbb{x}) \cong \begin{pmatrix}
 \dfrac{1}{\sqrt{y_1 y_3}} + \sqrt{\dfrac{y_1}{y_3}} & \sqrt{y_1} \\
 \sqrt{y_1} & \sqrt{y_1 y_3}
 \end{pmatrix} .
\end{gather*}

The Poisson algebra~\eqref{def_Poisson} for the $y$-variable with~\eqref{B_torus} proves the following (see~\cite{WGold97a}).

\begin{Proposition} \label{prop:Poisson_torus} From the embedding~$\iota$~\eqref{xyz_y}, we get the Poisson relations
 \begin{gather*}
 \{ x ,b\} = \{ y , b\} = \{ z , b\} = 0,
 \end{gather*}
 and
 \begin{gather*}
 \{ x, y\} = \frac{1}{2} x y -z = \frac{1}{2}\ \frac{\partial W_{\text{\rm Mar}}}{\partial z} , \\
 \{ y , z\} = \frac{1}{2} y z -x = \frac{1}{2}\ \frac{\partial W_{\text{\rm Mar}}}{\partial x} , \\
 \{ z , x\} = \frac{1}{2} z x - y = \frac{1}{2}\ \frac{\partial W_{\text{\rm Mar}}}{\partial y} ,
 \end{gather*}
 where $W_{\text{\rm Mar}}$ is defined in~\eqref{W_torus}.
\end{Proposition}

\subsection{Automorphism and braid group}

The automorphism group of $\mathcal{X}(\Sigma_{1,1})$ is generated by $\mathrm{PGL}(2;\mathbb{Z})$ with two sign changing gene\-ra\-tors~\cite{WGold03a, RHorow75a} (see also~\cite{GoldmNeuma05a}).

We shall construct the $\mathrm{PGL}(2;\mathbb{Z})$ action using the cluster mutations (see~\cite{HikamiRInoue12a,NagaTeraYama11a} for the $\mathrm{PSL}(2;\mathbb{Z})$ actions). We denote the generators of~$\mathrm{PSL}(2;\mathbb{Z})$ by
\begin{gather*} 
 \mathsf{R}= \begin{pmatrix}
 1 & 1 \\
 0 & 1
 \end{pmatrix},
 \qquad
 \mathsf{L} = \begin{pmatrix}
 1 & 0 \\
 1 & 1
 \end{pmatrix} .
\end{gather*}
These with
\begin{gather*} 
 \mathsf{M}=
 \begin{pmatrix}
 1 & 0 \\
 0 & -1
 \end{pmatrix} ,
\end{gather*}
generate $\mathrm{PGL}(2;\mathbb{Z})$. Those satisfy
\begin{gather} \label{def_PSL}
 \mathsf{S}^2= (\mathsf{S} \circ \mathsf{R})^3 = {\rm id}, \\
 \label{def_PGL}
 \mathsf{M}^2 = ( \mathsf{M} \circ \mathsf{R} )^2 = ( \mathsf{M} \circ \mathsf{L} )^2 = {\rm id},
\end{gather}
where $\mathsf{S}=\mathsf{R} \circ \mathsf{L}^{-1} \circ \mathsf{R}=\left( \begin{smallmatrix} 0 & 1 \\ -1 & 0 \end{smallmatrix}\right)$.

\begin{Proposition} The $\mathrm{PGL}(2;\mathbb{Z})$ actions on $\mathcal{A}(Q_{\text{\rm Mar}})$ are realized by
 \begin{gather} \label{RL_torus}
 \mathsf{R} = \sigma_{1,3} \mu_1, \qquad \mathsf{L} = \sigma_{2,3} \mu_2 , \\
 \mathsf{M}= \varsigma \mu_3 .\nonumber
 \end{gather}
 Here~$\sigma_{i,j}$ is a permutation of labels of vertices~$i$ and~$j$, and $\varsigma$ is defined in~\eqref{inversion_cluster}.
\end{Proposition}

See that the actions on the $y$-variables~\eqref{RL_torus} are explicitly written as
\begin{gather*}
 \mathsf{R}(\boldsymbol{y}) = \begin{pmatrix}
 y_3 \big( 1+{y_1}^{-1} \big)^{-2} \\
 y_2 ( 1+{y_1} )^{2} \\
 {y_1}^{-1}
 \end{pmatrix},
 \qquad
 \mathsf{L}(\boldsymbol{y}) = \begin{pmatrix}
 y_1 \big( 1+ {y_2}^{-1} \big)^{-2} \\
 y_3 ( 1+y_2 )^2 \\
 {y_2}^{-1}
 \end{pmatrix} , \\
 \mathsf{M}(\boldsymbol{y}) = \begin{pmatrix}
 {y_1}^{-1} (1+y_3t)^{-2} \\
 {y_2}^{-1} \big( 1+ {y_3}^{-1} \big)^2 \\
 y_3
 \end{pmatrix} .
\end{gather*}
We also see that the exchange matrix~\eqref{B_torus} is invariant, $\mathsf{R}(\mathbf{B}) =\mathsf{L}(\mathbf{B})=\mathbf{B}$, while $\mathsf{M}(\mathbf{B})= - \mathbf{B}$. Then we can check that the braid relation holds,
\begin{gather*}
 \mathsf{R} \circ \mathsf{L}^{-1} \circ \mathsf{R} = \mathsf{L}^{-1} \circ \mathsf{R} \circ \mathsf{L}^{-1}\colon \ \begin{pmatrix}
 y_1 \\ y_2 \\ y_3
 \end{pmatrix}
 \mapsto
 \begin{pmatrix}
 y_2 \big( 1+{y_3}^{-1} \big)^{-2} \\
 y_1 ( 1+y_3 )^2 \\
 {y_3}^{-1}
 \end{pmatrix} ,
\end{gather*}
and that the cyclic permutation on the $y$-variables is given by
\begin{gather*}
 \mathsf{L}^{-1} \circ \mathsf{R}\colon \
 \begin{pmatrix}
 y_1 \\ y_2 \\ y_3
 \end{pmatrix}
 \mapsto
 \begin{pmatrix}
 y_3 \\ y_1 \\ y_2
 \end{pmatrix} .
\end{gather*}
These prove~\eqref{def_PSL}, and it is also straightforward to see~\eqref{def_PGL}.

These cluster mutation actions~\eqref{RL_torus} induce the $\mathrm{PGL}(2;\mathbb{Z})$ action on $\mathcal{X}(\Sigma_{1,1})$ via Theorem~\ref{thm:torus_phi}.

\begin{Proposition} The actions $\mathsf{R}$, $\mathsf{L}$, and $\mathsf{M}$
 on $\mathcal{X}(\Sigma_{1,1})$ are read respectively as
 \begin{gather*} 
 \mathsf{R} \colon \ b \mapsto b, \qquad
 \begin{pmatrix}
 x \\ y \\ z
 \end{pmatrix}
 \mapsto
 \begin{pmatrix}
 x \\
 z \\
 x z - y
 \end{pmatrix} , \\
 \mathsf{L}\colon \ b \mapsto b, \qquad
 \begin{pmatrix}
 x \\ y \\ z
 \end{pmatrix}
 \mapsto
 \begin{pmatrix}
 z \\ y \\ y z - x
 \end{pmatrix} ,\\
 \mathsf{M}\colon \ b \mapsto b,
 \qquad
 \begin{pmatrix}
 x \\ y \\ z
 \end{pmatrix}
 \mapsto
 \begin{pmatrix}
 x \\ y \\ x y - z
 \end{pmatrix} .
 \end{gather*}
\end{Proposition}
These were given in~\cite{WGold03a}. We should note that these $\mathrm{PGL}(2;\mathbb{Z})$ actions preserve the Poisson structure in Proposition~\ref{prop:Poisson_torus}.

\section{Character varieties of 4-punctured sphere}\label{sec:sphere}

\subsection{Character variety and cubic surface}
\begin{figure}[htbp] \centering
 \begin{minipage}{0.4\linewidth}
 \centering
 \includegraphics[scale=1]{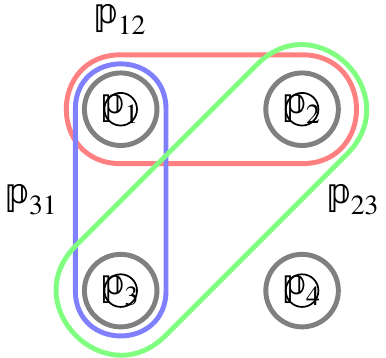}
 \end{minipage}
 \qquad
 \begin{minipage}{0.4\linewidth}
 \centering
 \includegraphics[scale=1]{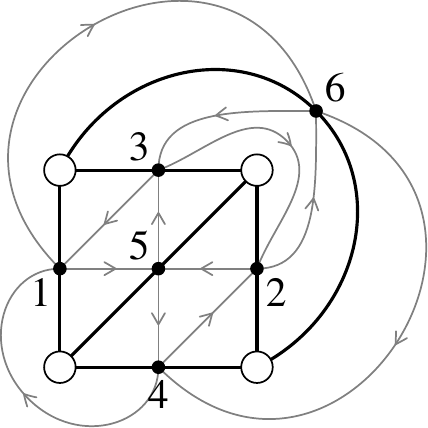}
 \end{minipage}
\caption{Closed paths on the $4$-punctured sphere are depicted in the left. An ideal triangulation of~$\Sigma_{0,4}$ is in the right, where white circles are nothing but the punctures in the left figure. Therein also depicted in gray is a quiver $Q_{\text{oct}}$.} \label{fig:sphere}
\end{figure}

On the $4$-punctured sphere $\Sigma_{0,4}$, we denote $\mathbb{p}_1$, $\mathbb{p}_2$, $\mathbb{p}_3$, $\mathbb{p}_4$ as closed paths encircling each puncture. We also use $\mathbb{p}_{12}$, $\mathbb{p}_{23}$, $\mathbb{p}_{31}$, in which $\mathbb{p}_{ab}$ is a loop around two punctures $a$ and $b$. See Fig.~\ref{fig:sphere}.

We shall study the $\mathrm{SL}(2;\mathbb{C})$ characters
\begin{gather*}
 p_{ab} = \Tr \rho(\mathbb{p}_{ab}) , \qquad p_i = \Tr \rho(\mathbb{p}_i) .
\end{gather*}
Well known since Fricke is that the character variety of $\Sigma_{0,4}$ is a hypersurface consisting of $(p_{12}, p_{23}, p_{31}, p_1, p_2 , p_3 , p_4)\in \mathbb{C}^7$ satisfying
\begin{gather} \label{W_is_0}
 W(p_{12},p_{23}, p_{31}; p_1, p_2, p_3, p_4)=0 ,
\end{gather}
where
\begin{gather}
 W(p_{12},p_{23}, p_{31}; p_1, p_2, p_3, p_4) = p_{12} p_{23} p_{31} - \bigl[ {p_{12}}^2 + {p_{23}}^2 + {p_{31}}^2 + (p_1 p_2 + p_3 p_4 ) p_{12} \nonumber\\
 \qquad{} + (p_1 p_3 + p_2 p_4 ) p_{31} + (p_1 p_4 + p_2 p_3 ) p_{23} + {p_1}^2+ {p_2}^2 + {p_3}^2+ {p_4}^2 + p_1 p_2 p_3 p_4 -4 \bigr] . \label{W_4puncture}
\end{gather}
We set
\begin{gather*}
 \mathcal{X}(\Sigma_{0,4}) = \big\{ (p_{12}, p_{31}, p_{23}) \in \mathbb{C}^3 \,|\, W(p_{12}, p_{31}, p_{23}; p_1, p_2, p_3, p_4)=0 \big\} ,
\end{gather*}
for a fixed set of $p_i$'s. Shown~\cite{Iwasaki02a} was that it has singular points when
\begin{gather}
 \prod_{i=1}^4 \big( {p_i}^2 -4 \big)\nonumber \\
 \times
 \Biggl\{ \prod_{ \substack{ \varepsilon_i=\pm 1 \\ \varepsilon_1 \varepsilon_2 \varepsilon_3=1 }}\!\!
 ( \varepsilon_1 p_1 + \varepsilon_2 p_2 + \varepsilon_3 p_3 +p_4) -( p_1 p_4-p_2 p_3 )( p_2 p_4-p_1 p_3) ( p_3 p_4-p_1 p_2) \Biggr\}=0 .\!\!\!\!\label{singular_cond}
\end{gather}

\subsection[Character varieties and $y$-variables]{Character varieties and $\boldsymbol{y}$-variables}

A triangulation of the 4-punctured sphere $\Sigma_{0,4}$ is given in Fig.~\ref{fig:sphere}. Therein we assign a vertex of quiver to each edge of the triangles, and give an anti-clockwise orientation to every dual face as depicted in gray lines. As a result, we obtain an octahedronic quiver $Q_{\text{oct}}$ as follows~\cite{HikamiRInoue12a}
\begin{gather} \label{octa_quiver}
 Q_{\text{oct}} =
 \raisebox{-17mm}{ \centering \includegraphics[scale=0.9]{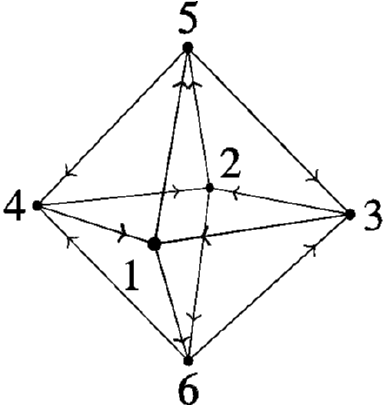} }
\end{gather}
The exchange matrix $\mathbf{B}_{\text{oct}}$ for $Q_{\text{oct}}$ is
\begin{gather} \label{B_sphere}
 \mathbf{B}_{\text{oct}} =
 \begin{pmatrix}
 0 & 0 & -1 & -1 & 1 & 1 \\
 0 & 0 & -1 & -1 & 1 & 1 \\
 1 & 1 & 0 & 0 & -1 & -1 \\
 1 & 1 & 0 & 0 & -1 & -1 \\
 -1 & -1 & 1 & 1 & 0 & 0 \\
 -1 & -1 & 1 & 1 & 0 & 0
 \end{pmatrix} .
\end{gather}

We employ Bonahon--Wong's trace map~\cite{BonaWong11a,TQLe16a} as well in this case of the $4$-punctured sphere. We have four copies of the triangle algebra~\eqref{triangle_algebra} which is assigned to each face of
Fig.~\ref{fig:sphere}, and the $y$-variables are given from the generators of the triangle algebras. By use of the $y$-variables, we obtain the following realization.

\begin{Theorem} \label{thm:map_sphere} We have an embedding
 \begin{gather*}
 \iota\colon \ \mathcal{X}(\Sigma_{0,4}) \to \mathcal{A}(Q_{\text{\rm oct}}) ,
 \end{gather*}
 defined by
 \begin{alignat}{3}
& \iota({p}_{1})= \sqrt{y_1 y_3 y_6} + \frac{1}{\sqrt{y_1 y_3 y_6}} , \qquad && \iota({p}_2)= \sqrt{y_2 y_3 y_5} + \frac{1}{\sqrt{y_2 y_3 y_5}} ,&\nonumber \\
& \iota({p}_3)= \sqrt{y_1 y_4 y_5} + \frac{1}{\sqrt{y_1 y_4 y_5}} ,\qquad && \iota( {p}_4 )= \sqrt{y_2 y_4 y_6} + \frac{1}{\sqrt{y_2 y_4 y_6}} ,& \label{b_from_y}
 \end{alignat}
 and
 \begin{gather}
 \iota(p_{12})= \sqrt{y_1 y_2 y_5 y_6} + \sqrt{\frac{y_1 y_2 y_6}{y_5}} + \sqrt{\frac{y_1 y_2 y_5}{y_6}}\nonumber \\
 \hphantom{\iota(p_{12})=}{} + \sqrt{\frac{y_1 y_2}{y_5 y_6}} + \sqrt{\frac{y_2}{y_1 y_5 y_6}} + \sqrt{\frac{y_1}{y_2 y_5 y_6}} + \frac{1}{ \sqrt{y_1 y_2 y_5 y_6} },\nonumber\\
\iota(p_{31})= \sqrt{y_3 y_4 y_5 y_6} + \sqrt{\frac{y_4 y_5 y_6}{y_3}} + \sqrt{\frac{y_3 y_5 y_6}{y_4}}\nonumber \\
 \hphantom{\iota(p_{31})=}{} + \sqrt{\frac{y_5 y_6}{y_3 y_4}} + \sqrt{\frac{y_6}{y_3 y_4 y_5}} + \sqrt{\frac{y_5}{y_3 y_4 y_6}} + \frac{1}{ \sqrt{y_3 y_4 y_5 y_6} },\nonumber\\
\iota(p_{23})= \sqrt{y_1 y_2 y_3 y_4} + \sqrt{\frac{y_2 y_3 y_4}{y_1}} + \sqrt{\frac{y_1 y_3 y_4}{y_2}}\nonumber \\
\hphantom{\iota(p_{23})=}{} + \sqrt{\frac{y_3 y_4}{y_1 y_2}} + \sqrt{\frac{y_4}{y_1 y_2 y_3}} + \sqrt{\frac{y_3}{y_1 y_2 y_4}} + \frac{1}{ \sqrt{y_1 y_2 y_3 y_4} }. \label{x_from_y}
 \end{gather}
\end{Theorem}

We can check from tedious computations that~\eqref{b_from_y} and~\eqref{x_from_y} fulfill~\eqref{W_is_0}. We note that slightly different parametrizations were given in~\cite{ChekhMazzo10a,ComaGabeTesc15a, GaioMoorNeit13a}.
The difference is crucial for the following studies of the automorphism, the confluences of the punctures, and the quantization from the view point of the cluster algebra associated to the quiver $Q_{\text{oct}}$. We remark that it suggests that we have up to conjugation
\begin{gather*}
 \rho(\mathbb{p}_{12}) \cong
 \begin{pmatrix}
 \dfrac{1+y_2}{ \sqrt{y_1 y_2 y_5 y_6} } & \dfrac{1+y_2 + y_2 y_6}{\sqrt{y_2 y_5 y_6}} \vspace{1mm}\\
 \dfrac{1+y_2 + y_2 y_5}{\sqrt{y_2 y_5 y_6}} & \sqrt{\dfrac{y_1}{y_2 y_5 y_6}}( 1+y_2 + y_2 y_5 + y_2 y_6 + y_2 y_5 y_6)
 \end{pmatrix} .
\end{gather*}
It is noted that, under the map~\eqref{b_from_y}, the condition~\eqref{singular_cond} for singularity is factorized as
\begin{gather*}
(y_1-y_2) ( y_3 - y_4) ( y_5-y_6 ) ( 1-y_1 y_2 y_3 y_4 y_5 y_6 )\cdot \prod_{(i,j,k) \in \triangle} ( 1-y_i y_j y_k ) =0 ,
\end{gather*}
where $\triangle$ denotes $8$ faces of the octahedral quiver~\eqref{octa_quiver}, $\{ (1{,}3{,}5), (2{,}3{,}5), (1{,}4{,}5), (2{,}4{,}5), (1{,}3{,}6)$, $(1,4,6), (2,3,6), (2,4,6)\}$.

The Poisson structure~\eqref{def_Poisson} of the cluster $y$-variables with~\eqref{B_sphere} gives the following Goldman Poisson algebra of $\mathcal{X}(\Sigma_{0,4})$ (see, e.g.,~\cite{ChekhMazzo10a,ChekMazzRubt15a,NekrRoslShat11a}).

\begin{Proposition} \label{prop:Poisson_x_b} We have
 \begin{gather} \label{b_b_Poisson}
\{ p_i, p_j \} = 0 ,\qquad \{ p_{12} , p_j \} = \{ p_{31} , p_j \} =\{ p_{23} , p_j\} = 0 ,
 \end{gather}
 and
 \begin{gather}
\{ p_{12}, p_{31} \} = p_{12} p_{31} - 2 p_{23} - ( p_1 p_4 + p_2 p_3 ) = \frac{\partial W}{\partial p_{23}} ,\nonumber \\
\{ p_{31} , p_{23}\} = p_{31} p_{23} - 2 p_{12} - ( p_1 p_2 + p_3 p_4 ) = \frac{\partial W}{\partial p_{12}} ,\nonumber \\
\{ p_{23} , p_{12} \} = p_{12} p_{23} - 2 p_{31} - ( p_1 p_3 + p_2 p_4 ) = \frac{\partial W}{\partial p_{31}} , \label{x_x_Poisson}
 \end{gather}
 where $W$ is given in~\eqref{W_is_0}.
\end{Proposition}

\subsection{Automorphism and braid group}
We study the $\mathrm{PGL}(2;\mathbb{Z})$ actions on the cluster algebra $\mathcal{A}(Q_{\text{oct}})$. As in the case of the once-punctured torus, we introduce a set of cluster mutations as the automorphism of the triangulations (see~\cite{HikamiRInoue12a} for $\mathrm{PSL}(2;\mathbb{Z})$).

\begin{Proposition} The $\mathrm{PGL}(2;\mathbb{Z})$ action on $\mathcal{A}(Q_{\text{\rm oct}})$ are generated by
 \begin{gather} \label{RL_mutation}
\mathsf{R} = \sigma_{5,6} \sigma_{1,5} \sigma_{2,6} \mu_1 \mu_2 ,\qquad \mathsf{L} = \sigma_{5,6} \sigma_{3,5} \sigma_{4,6} \mu_3 \mu_4 ,\qquad \mathsf{M}= \varsigma \sigma_{5,6} \mu_5 \mu_6 ,
\end{gather}
where $\sigma_{i,j}$ is a permutation of $i$ and $j$, and $\varsigma$ is defined in~\eqref{inversion_cluster}.
\end{Proposition}

We see from~\eqref{mutation_B} that the exchange matrix~\eqref{B_sphere} is invariant under $\mathrm{PSL}(2;\mathbb{Z})$, $\mathsf{R}(\mathbf{B})=\mathsf{L}(\mathbf{B})=\mathbf{B}$, while $\mathsf{M}(\mathbf{B})=-\mathbf{B}$. It can be also seen that the actions on $y$-variables are explicitly written as
 \begin{gather*}
 \mathsf{R}(\boldsymbol{y}) =
 \begin{pmatrix}
 y_5 \big( 1+ {y_1}^{-1} \big)^{-1}
 \big( 1+{y_2}^{-1} \big)^{-1} \\
 y_6 \big( 1+ {y_1}^{-1} \big)^{-1}
 \big( 1+{y_2}^{-1} \big)^{-1} \\
 y_3 ( 1+y_1 ) ( 1+y_2 ) \\
 y_4 ( 1+y_1 ) ( 1+y_2 ) \\
 {y_2}^{-1} \\
 {y_1}^{-1}
 \end{pmatrix} , \qquad
 \mathsf{L}(\boldsymbol{y}) =
 \begin{pmatrix}
 y_1 \big( 1+ {y_3}^{-1} \big)^{-1}
 \big( 1+{y_4}^{-1} \big)^{-1} \\
 y_2 \big( 1+ {y_3}^{-1} \big)^{-1}
 \big( 1+{y_4}^{-1} \big)^{-1} \\
 y_5 ( 1+y_3 ) ( 1+y_4 ) \\
 y_6 ( 1+y_3 ) ( 1+y_4 ) \\
 {y_4}^{-1} \\
 {y_3}^{-1} \\
 \end{pmatrix} , \\
 \mathsf{M}(\boldsymbol{y}) =
 \begin{pmatrix}
 {y_1}^{-1} ( 1+y_5 )^{-1} ( 1+y_6 )^{-1} \\
 {y_2}^{-1} ( 1+y_5 )^{-1} ( 1+y_6 )^{-1} \\
 {y_3}^{-1} \big( 1+{y_5}^{-1} \big) \big( 1+{y_6}^{-1} \big) \\
 {y_4}^{-1} \big( 1+{y_5}^{-1} \big) \big(1+{y_6}^{-1} \big) \\
 y_6 \\
 y_5
 \end{pmatrix} .
 \end{gather*}
 Using these expressions, we find the braid relation,
 \begin{gather*}
 \mathsf{R} \circ \mathsf{L}^{-1} \circ \mathsf{R} = \mathsf{L}^{-1} \circ \mathsf{R} \circ \mathsf{L}^{-1}\colon \
 \boldsymbol{y}
 \mapsto
 \begin{pmatrix}
 y_3 \big( 1+{y_5}^{-1} \big)^{-1} \big( 1+{y_6}^{-1} \big)^{-1} \\
 y_4 \big( 1+{y_5}^{-1} \big)^{-1} \big( 1+{y_6}^{-1} \big)^{-1} \\
 y_1 ( 1+y_5 ) ( 1+y_6 ) \\
 y_2 ( 1+y_5 ) ( 1+y_6 ) \\
 {y_6}^{-1} \\
 {y_5}^{-1}
 \end{pmatrix} ,
 \end{gather*}
 and a cyclic permutation on $y$-variables is given by
 \begin{gather*}
 \mathsf{L}^{-1} \circ \mathsf{R}\colon \
 \boldsymbol{y} \mapsto
 \left( \begin{matrix}
 y_5 \\ y_6 \\ y_1 \\ y_2 \\ y_3 \\ y_4
 \end{matrix} \right) ,
 \end{gather*}
 which proves~\eqref{def_PSL}. We can also check~\eqref{def_PGL} on $\mathcal{A}(Q_{\text{oct}})$. It should be stressed that the $\mathrm{PGL}(2;\mathbb{Z})$ action preserves the Poisson algebra~\eqref{b_b_Poisson}--\eqref{x_x_Poisson}.

The map $\iota\colon \mathcal{X}(\Sigma_{0,4}) \to\mathcal{A}(Q_{\text{oct}})$ induces the following actions on the character varieties.

\begin{Proposition} We have the $\mathrm{PGL}(2;\mathbb{Z})$ actions on $\mathcal{X}(\Sigma_{0,4})$ as
 \begin{alignat*}{3}
 & \mathsf{R} \colon \ \left( \begin{matrix}
 {p}_1 \\
 {p}_2 \\
 {p}_3 \\
 {p}_4
 \end{matrix} \right) \mapsto \left( \begin{matrix}
 {p}_2 \\
 {p}_1 \\
 {p}_3 \\
 {p}_4
 \end{matrix}
 \right),
 \qquad &&
 \left( \begin{matrix}
 p_{12} \\
 p_{31} \\
 p_{23}
 \end{matrix}
 \right)
 \mapsto
 \left( \begin{matrix}
 p_{12} \\
 p_{23} \\
 p_{23} p_{12} - p_{31} - p_1 p_3 - p_2 p_4
 \end{matrix} \right) ,& \\
 & \mathsf{L}\colon \ \left( \begin{matrix}
 {p}_1 \\
 {p}_2 \\
 {p}_3 \\
 {p}_4
 \end{matrix}
 \right)
 \mapsto
 \left( \begin{matrix}
 {p}_3 \\
 {p}_2 \\
 {p}_1 \\
 {p}_4
 \end{matrix}
 \right),
 \qquad &&
 \left( \begin{matrix}
 p_{12} \\
 p_{31} \\
 p_{23}
 \end{matrix}
 \right)
 \mapsto \left( \begin{matrix}
 p_{23} \\
 p_{31} \\
 p_{31} p_{23} - p_{12} - p_1 p_2 - p_3 p_4
 \end{matrix}
 \right) , & \\
 & \mathsf{M} \colon \
 \begin{pmatrix}
 p_1 \\ p_2 \\ p_3 \\ p_4
 \end{pmatrix}
 \mapsto
 \begin{pmatrix}
 p_1 \\ p_2 \\ p_3 \\ p_4
 \end{pmatrix} ,
 \qquad &&
 \begin{pmatrix}
 p_{12} \\ p_{31} \\ p_{23}
 \end{pmatrix}
 \mapsto
 \begin{pmatrix}
 p_{12} \\ p_{31} \\
 p_{12} p_{31} - p_{23}
 - p_1 p_4 - p_2 p_3
 \end{pmatrix} .&
\end{alignat*}
\end{Proposition}

The braid group action generated by $\mathsf{R}$ and $\mathsf{L}$ on $\mathcal{X}(\Sigma_{0,4})$ is related to an analytic continuation of the Painlev{\'e} VI equation~\cite{Iwasaki02a, Iwasaki03a}. We should stress that the $\mathrm{PSL}(2;\mathbb{Z})$ braid group action is realized by mutations of the cluster algebra, while it was studied using a ``generalization'' of cluster algebra in~\cite{ChekhShapi14a}.

\section{Confluence of punctures on sphere}\label{sec:confluence}
 We study degenerations of the character varieties $\mathcal{X}(\Sigma_{0,4})$ of the $4$-punctured sphere. As the Painlev{\'e} VI equations are related to the monodromy preserving problem with four punctures (see, e.g.,~\cite{MJimb82a}), we have in mind confluences of the Painlev{\'e} equations which is classified up-to-dately as~\cite{OhyamOkumu06a}
\begin{equation} \label{Painleve_confluence}
 \begin{tikzcd}
 & &
 P_{\mathrm{III}} \arrow[r] \arrow[dr] &
 P_{\mathrm{III}}^{D_7} \arrow[r] \arrow[dr]&
 P_{\mathrm{III}}^{D_8}
 \\
 P_{\mathrm{VI}}
 \arrow[r] &
 P_{\mathrm{V}}
 \arrow[ur] \arrow[r] \arrow[dr] &
 P_{\mathrm{V}}^{\text{deg}} \arrow[ur] \arrow{rd}&
 P_{\mathrm{II}}^{\text{JM}} \arrow[r] &
 P_{\mathrm{I}}.
 \\
 & &
 P_{\mathrm{IV}} \arrow[ur] \arrow[r] &
 P_{\mathrm{II}}^{\text{FN}} \arrow[ur] &
 \end{tikzcd}
\end{equation}
Correspondingly the moduli space~\eqref{W_is_0} could be degenerated, and $10$ families of moduli spaces as generalized monodromy data were proposed in~\cite{vdPutSaito09a}. Their geometrical aspects was studied in~\cite{ChekMazzRubt15a} from a viewpoint of the Teichm{\"u}ller theory of a punctured Riemann surface.

\begin{figure} \centering
 \begin{tikzpicture}
 [auto,
 myblock/.style={rectangle, draw=black , thick, rounded corners,
 align=center },
 mytitb/.style={fill=white},
 myarrow/.style={decoration={markings, mark=at position .83 with
 {\arrow[scale=2.5, >=stealth
 ]{>}}}, postaction={decorate}}
 ]
 \draw (0,0) node[myblock] (P1111) {
 \makecell{
 \\
 \raisebox{-4mm}{
 \includegraphics[scale=.71]{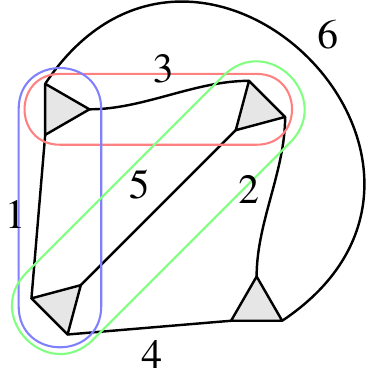}
 }
 \quad
 \scalebox{.71}{
 \begin{tikzpicture}
 \coordinate (B1) at (0,1.6);
 \coordinate (B2) at (1.6,1.6);
 \coordinate (B3) at (0,0);
 \coordinate (B4) at (1.6,0);
 \coordinate (BC) at (0.4, 0.8);
 \coordinate (BD) at (1.2, 0.8);
 \node [above left] at (B1) {$4$};
 \node [above right] at (B2) {$2$};
 \node [below left] at (B3) {$1$};
 \node [below right] at (B4) {$3$};
 \node [left ] at (BC) {$5$};
 \node [right ] at (BD) {$6$};
 \draw[->-=.9] (B3) -- (BD);
 \draw[->-=.7] (BD) -- (B1);
 \draw[line width=7pt, draw=white] (B2) -- (BC);
 \draw[->-=.8] (B2) -- (BC);
 \draw[line width=7pt, draw=white] (BC) -- (B4);
 \draw[->-=.7] (BC) -- (B4);
 \filldraw (B1) circle[radius=1.2pt];
 \filldraw (B2) circle[radius=1.2pt];
 \filldraw (B3) circle[radius=1.2pt];
 \filldraw (B4) circle[radius=1.2pt];
 \filldraw (BC) circle[radius=1.2pt];
 \filldraw (BD) circle[radius=1.2pt];
 \draw[->-=.7] (B1) -- (B2);
 \draw[->-=.8] (B1) -- (B3);
 \draw[->-=.8] (B4) -- (B2);
 \draw[->-=.7] (B4) -- (B3);
 \draw[->-=.7] (B2) -- (BD);
 \draw[->-=.7] (BD) -- (B4);
 \draw[->-=.8] (B3) -- (BC);
 \draw[->-=.7] (BC) -- (B1);
 \end{tikzpicture}
 }
 }
 } ;
 \node[mytitb] at (P1111.north) {$\mathcal{P}_{(1,1,1,1)}$};
 \draw(8,-1) node[myblock] (P211) {
 \makecell{
 \\
 \raisebox{-5mm}{
 \includegraphics[scale=.71]{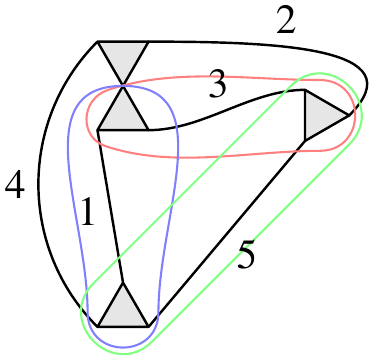}
 }
 \quad
 \scalebox{.71}{
 \begin{tikzpicture}
 \coordinate (B1) at (0,1.6);
 \coordinate (B2) at (1.6,1.6);
 \coordinate (B3) at (0,0);
 \coordinate (B4) at (1.6,0);
 \coordinate (BC) at (0.8, 0.8);
 \node [above left] at (B1) {$4$};
 \node [above right] at (B2) {$2$};
 \node [below left] at (B3) {$1$};
 \node [below right] at (B4) {$3$};
 \node [above ] at (BC) {$5$};
 \filldraw (B1) circle[radius=1.2pt];
 \filldraw (B2) circle[radius=1.2pt];
 \filldraw (B3) circle[radius=1.2pt];
 \filldraw (B4) circle[radius=1.2pt];
 \filldraw (BC) circle[radius=1.2pt];
 \draw[->-=.7] (B1) -- (B2);
 \draw[->-=.7] (B1) -- (B3);
 \draw[->-=.7] (B4) -- (B2);
 \draw[->-=.7] (B4) -- (B3);
 \draw[->-=.7] (B2) -- (BC);
 \draw[->-=.7] (B3) -- (BC);
 \draw[->-=.7] (BC) -- (B1);
 \draw[->-=.7] (BC) -- (B4);
 \end{tikzpicture}
 }
 }
 };
 \node[mytitb] at (P211.north) {$\mathcal{P}_{(2,1,1)}$};
 \draw(0,-7) node[myblock] (P31) {
 \makecell{
 \\[-3mm]
 \includegraphics[scale=.71]{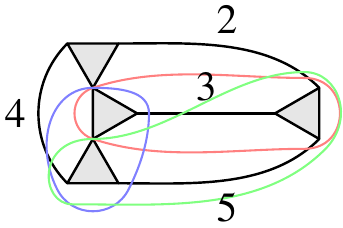}
 \\
 \scalebox{.71}{
 \begin{tikzpicture}
 \coordinate (B1) at (0,1.);
 \coordinate (B2) at (1.,1.);
 \coordinate (B3) at (0,0);
 \coordinate (B4) at (1.,0);
 \node [above left] at (B1) {$4$};
 \node [above right] at (B2) {$2$};
 \node [below left] at (B3) {$5$};
 \node [below right] at (B4) {$3$};
 \filldraw (B1) circle[radius=1.2pt];
 \filldraw (B2) circle[radius=1.2pt];
 \filldraw (B3) circle[radius=1.2pt];
 \filldraw (B4) circle[radius=1.2pt];
 \draw[->-=.7] (B1) -- (B2);
 \draw[->-=.7] (B3) -- (B1);
 \draw[->-=.7] (B4) -- (B2);
 \draw[->-=.7] (B3) -- (B4);
 \draw[->-=.7] (B2) -- (B3);
 \end{tikzpicture}
 }
 }
 };
 \node[mytitb] at (P31.north) {$\mathcal{P}_{(3,1)}$};
 \draw(5,-7) node[myblock] (P211deg) {
 \makecell{
 \\
 \includegraphics[scale=.71]{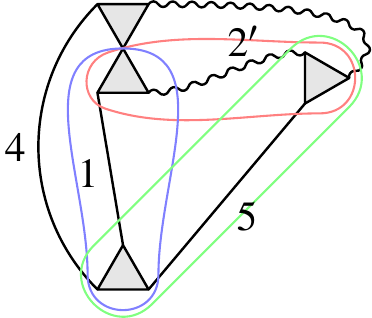}
 \\
 \scalebox{.71}{
 \begin{tikzpicture}
 \coordinate (B1) at (0,1.);
 \coordinate (B2) at (1.,1.);
 \coordinate (B3) at (0,0);
 \coordinate (B4) at (1.,0);
 \node [above left] at (B1) {$5$};
 \node [above right] at (B2) {$4$};
 \node [below left] at (B3) {$1$};
 \node [below right] at (B4) {$2^\prime$};
 \filldraw (B1) circle[radius=1.2pt];
 \filldraw (B2) circle[radius=1.2pt];
 \filldraw (B3) circle[radius=1.2pt];
 \filldraw (B4) circle[radius=1.2pt];
 \draw[->-=.7] (B1) -- (B2);
 \draw[->-=.7] (B3) -- (B1);
 \draw[->-=.7] (B2) -- (B4);
 \draw[->-=.7] (B4) -- (B3);
 \draw[->-=.7] (B2) -- (B3);
 \end{tikzpicture}
 }
 }
 };
 \node[mytitb] at (P211deg.north) {$\mathcal{P}_{(2,1,1)}^{\text{deg}}$};
 \draw(10,-7) node[myblock] (P22) {
 \makecell{
 \\
 \includegraphics[scale=.71]{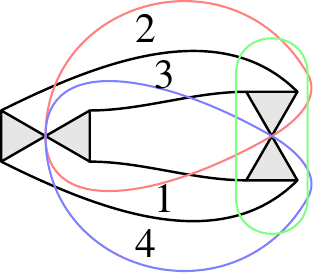}
 \\
 \scalebox{.71}{
 \begin{tikzpicture}
 \coordinate (B1) at (0,1.);
 \coordinate (B2) at (1.,1.);
 \coordinate (B3) at (0,0);
 \coordinate (B4) at (1.,0);
 \node [above left] at (B1) {$4$};
 \node [above right] at (B2) {$2$};
 \node [below left] at (B3) {$1$};
 \node [below right] at (B4) {$3$};
 \filldraw (B1) circle[radius=1.2pt];
 \filldraw (B2) circle[radius=1.2pt];
 \filldraw (B3) circle[radius=1.2pt];
 \filldraw (B4) circle[radius=1.2pt];
 \draw[->-=.7] (B1) -- (B2);
 \draw[->-=.7] (B1) -- (B3);
 \draw[->-=.7] (B4) -- (B2);
 \draw[->-=.7] (B4) -- (B3);
 \end{tikzpicture}
 }
 }
 } ;
 \node[mytitb] at (P22.north) {$\mathcal{P}_{(2,2)}$};
 \draw(3.5,-13) node[myblock] (P4A) {
 \makecell{
 \\
~
 \includegraphics[scale=.71]{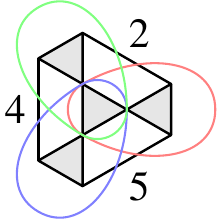}
 \\
 \scalebox{.71}{
 \begin{tikzpicture}
 \coordinate (B2) at (.5,{sqrt(3)/2});
 \coordinate (B3) at (0,0);
 \coordinate (B4) at (1.,0);
 \node [above right] at (B2) {$4$};
 \node [below left] at (B3) {$5$};
 \node [below right] at (B4) {$2$};
 \filldraw (B2) circle[radius=1.2pt];
 \filldraw (B3) circle[radius=1.2pt];
 \filldraw (B4) circle[radius=1.2pt];
 \draw[->-=.7] (B2) -- (B4);
 \draw[->-=.7] (B4) -- (B3);
 \draw[->-=.7] (B3) -- (B2);
 \end{tikzpicture}
 }
 }
 };
 \node[mytitb] at (P4A.north) {$\mathcal{P}_{(4)A}$};
 \draw(-1,-13) node[myblock] (P31B) {
 \makecell{
 \\
 \includegraphics[scale=.71]{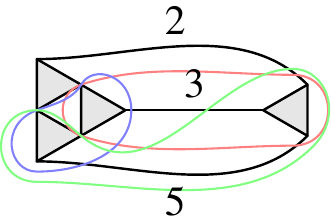}
 \\
 \scalebox{.71}{
 \begin{tikzpicture}
 \coordinate (B2) at (.5,{sqrt(3)/2});
 \coordinate (B3) at (0,0);
 \coordinate (B4) at (1.,0);
 \node [above right] at (B2) {$2$};
 \node [below left] at (B3) {$5$};
 \node [below right] at (B4) {$3$};
 \filldraw (B2) circle[radius=1.2pt];
 \filldraw (B3) circle[radius=1.2pt];
 \filldraw (B4) circle[radius=1.2pt];
 \draw[->-=.7] (B4) -- (B2);
 \draw[->-=.7] (B3) -- (B4);
 \draw[->-=.7] (B2) -- (B3);
 \end{tikzpicture}
 }
 }
 };
 \node[mytitb] at (P31B.north) {$\mathcal{P}_{(3,1)B}$};
 \draw(6.5,-13) node[myblock] (P4B) {
 \makecell{
 \\
 \includegraphics[scale=.71]{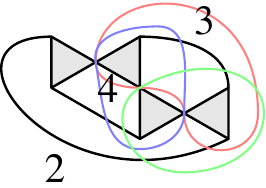}
 \\
 \scalebox{.71}{
 \begin{tikzpicture}
 \coordinate (B2) at (.5,{sqrt(3)/2});
 \coordinate (B3) at (0,0);
 \coordinate (B4) at (1.,0);
 \node [above right] at (B2) {$4$};
 \node [below left] at (B3) {$3$};
 \node [below right] at (B4) {$2$};
 \filldraw (B2) circle[radius=1.2pt];
 \filldraw (B3) circle[radius=1.2pt];
 \filldraw (B4) circle[radius=1.2pt];
 \draw[->-=.7] (B2) -- (B4);
 \draw[->-=.7] (B3) -- (B4);
 \end{tikzpicture}
 }
 }
 };
 \node[mytitb] at (P4B.north) {$\mathcal{P}_{(4)B}$};
 \draw(11,-13) node[myblock] (P22deg) {
 \makecell{
 \\
 \includegraphics[scale=.71]{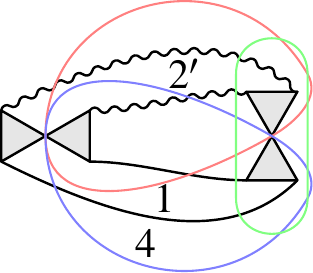}
 \\
 \scalebox{.71}{
 \begin{tikzpicture}
 \coordinate (B1) at (0,1);
 \coordinate (B24) at ({sqrt(3)/2},0.5);
 \coordinate (B3) at (0,0);
 \node [above left] at (B1) {$4$};
 \node [ right] at (B24) {$2^\prime$};
 \node [below left] at (B3) {$1$};
 \filldraw (B1) circle[radius=1.2pt];
 \filldraw (B24) circle[radius=1.2pt];
 \filldraw (B3) circle[radius=1.2pt];
 \draw[->-=.7] (B1) -- (B24);
 \draw[->-=.7] (B1) -- (B3);
 \draw[->-=.7] (B24) -- (B3);
 \end{tikzpicture}
 }
 }
 };
 \node[mytitb] at (P22deg.north) {$\mathcal{P}_{(2,2)}^{\text{deg}}$};
 \draw(3,-18) node[myblock] (P4C) {
 \makecell{
 \\
~~
 \includegraphics[scale=.71]{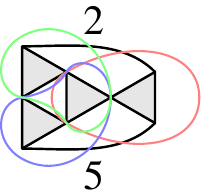}
 \\[1ex]
 \scalebox{.71}{
 \begin{tikzpicture}
 \coordinate (B3) at (0,0);
 \coordinate (B4) at (1.,0);
 \node [below left] at (B3) {$5$};
 \node [below right] at (B4) {$2$};
 \filldraw (B3) circle[radius=1.2pt];
 \filldraw (B4) circle[radius=1.2pt];
 \draw[->-=.7] (B4) -- (B3);
 \end{tikzpicture}
 }
 }
 };
 \node[mytitb] at (P4C.north) {$\mathcal{P}_{(4)C}$};
 \draw(9,-18) node[myblock] (P22degP) {
 \makecell{
 \\
 \includegraphics[scale=.71]{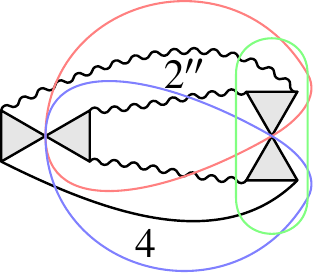}
 \\[1ex]
 \scalebox{.71}{
 \begin{tikzpicture}
 \coordinate (B3) at (0,0);
 \coordinate (B4) at (1.,0);
 \node [below left] at (B3) {$2^{\prime\prime}$};
 \node [below right] at (B4) {$4$};
 \filldraw (B3) circle[radius=1.2pt];
 \filldraw (B4) circle[radius=1.2pt];
 \draw[->-=.7] (B4) to [bend left=30] (B3);
 \draw[->-=.7] (B4) to [bend right=30] (B3);
 \end{tikzpicture}
 }
 }
 };
 \node[mytitb] at (P22degP.north) {$\mathcal{P}_{(2,2)}^{\text{deg}^2}$};
 \draw[myarrow] (P1111.east) [out=0, in=165] to (P211.west);
 \draw[myarrow] (P211.south west) to [bend right=30] (P31.north east);
 \draw[myarrow, dashed] (P211) --(P211deg);
 \draw[myarrow] (P211.south) -- (P22.north west);
 \draw[myarrow] (P31) [out=-80, in=120] to (P4A.north west);
 \draw[myarrow] (P31.south) [out=-90, in=70]to (P31B);
 \draw[myarrow] (P31.south east) [out=-70, in=120] to (P4B);
 \draw[myarrow] (P211deg.south west) [out=-120, in=50] to (P31B.north east);
 \draw[myarrow] (P211deg) [out=-60,in=120] to (P22deg);
 \draw[myarrow] (P22) [out=-110,in=70] to (P4B);
 \draw[myarrow, dashed] (P22) [out=-80,in=110] to (P22deg);
 \draw[myarrow] (P4A.south) [out=-100, in=70] to (P4C);
 \draw[myarrow] (P31B) [out=-70, in=110] to (P4C.north west);
 \draw[myarrow] (P4B) [out=-100, in=70] to (P4C.north east);
 \draw[myarrow, dashed] (P22deg.south) to [bend left=20] (P22degP.north east);
 \end{tikzpicture}

 \caption{Confluences of the punctures. The shaded triangles denote the punctures, and the colored curves are $\mathbb{p}_{ab}$ as in Fig.~\ref{fig:sphere}. The quiver~$Q_J$, as a degeneration of $Q_{\text{oct}}$, is also given for each $J$. We have two types of degenerations (solid and dashed arrows). } \label{fig:degenerate}
\end{figure}
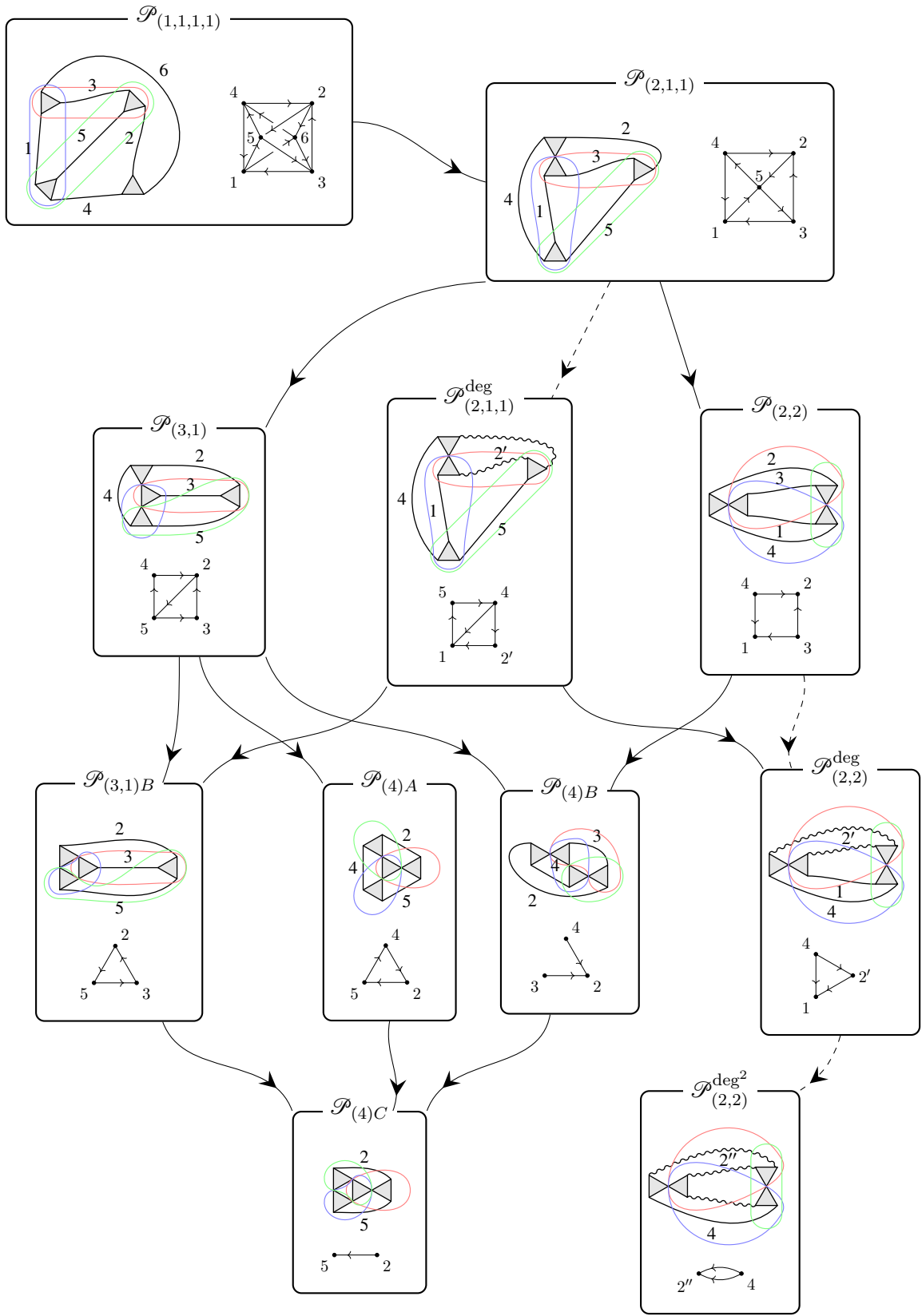

Motivated by~\cite{ChekMazzRubt15a,ChekMazzRubt17a}, we shall study confluences of the punctures from a cluster algebraic viewpoint of the moduli space. In our previous studies on geometric aspects of the cluster algebra~\cite{HikamiRInoue12a}, the cluster $y$-variables may be regarded as a~modulus of an ideal tetrahedron in 3-dimensional hyperbolic space $\mathbb{H}^3$. An ideal tetrahedron is that
all vertices are on boundary~$\partial \mathbb{H}^3$, and it is a truncated tetrahedron. Hence we denote each puncture as a triangle hereafter, and we propose degenerations of the punctures as $\mathcal{P}_J=\text{Puncture}_J$ depicted in Fig.~\ref{fig:degenerate}, where we have $11$ families of diagrams. Basically $J$ denotes a partition of $4$ as a confluence of the four punctures. With the exception of $\mathcal{P}_{(4)A}$, we see a one-to-one correspondence with the Painlev{\'e} confluence~\eqref{Painleve_confluence}. We will see later that the moduli spaces for $\mathcal{P}_{(4)A}$, $\mathcal{P}_{(4)B}$, and $\mathcal{P}_{(3,1)B}$ are in the same class. In our proposal of confluences, we introduce two types of degenerations; dashed arrows in Fig.~\ref{fig:degenerate} mean that one of the Poisson commuting parameters vanish. This can be realized by renormalizing and reorganizing two cluster $y$-variables, which are depicted in squiggly curves in the figure.

In Fig.~\ref{fig:degenerate} we also define the quivers $Q_J$ associated to the confluent punctures $\mathcal{P}_J$, which can be viewed as a degeneration of the octahedronic quiver $Q_{\text{oct}}$~\eqref{octa_quiver}. Those are finite-mutation-type quivers~\cite{FomiShapThur08a}. See~\cite{FeliShapTuma12a} for a relationship with the $\lambda$-length (see~\cite{Penner12}).

As degenerations of the character varieties~$\mathcal{X}(\Sigma_{0,4})$, we define
\begin{gather*}
 \mathcal{X}_J = \big\{( p_{12}, p_{31}, p_{23}) \in \mathbb{C}^3 \,|\, W_J(p_{12}, p_{31},p_{23}; p_1,\dots, p_{|J|}) =0 \big\} ,
\end{gather*}
where an explicit form of $W_J$ is given below. Here we have the Poisson structure such that $p_i$'s are Poisson commutative,
\begin{gather*}
\{ p_i , p_j\} =0, \qquad \{ p_{12} , p_i \}= \{ p_{31}, p_i \}= \{ p_{23} , p_i\} = 0 , \\
\{ p_{12} , p_{31}\} = \frac{\partial W_J}{\partial p_{23}} ,\qquad \{ p_{31} , p_{23} \} = \frac{\partial W_J}{\partial p_{12}} ,\qquad \{ p_{23} , p_{12} \} = \frac{\partial W_J}{\partial p_{31}} .
\end{gather*}
Here $J$ corresponds to $\mathcal{P}_J$ in Fig~\ref{fig:degenerate}. It is noted that $\mathcal{P}_{(1,1,1,1)}$ is for $\Sigma_{0,4}$ studied in the previous section. Our claim in this section is the following.
\begin{Theorem} \label{thm:degenerate} We have an algebra embedding
\begin{gather*}
\iota_J\colon \ \mathcal{X}_J \to \mathcal{A}(Q_J),
\end{gather*}
where $Q_J$ denotes a quiver which is a degeneration of $Q_{\text{\rm oct}}$ as in Fig.~{\rm \ref{fig:degenerate}}. Explicit forms of $\iota_J$ are given below for each $J$.
\end{Theorem}

In the following subsections, we demonstrate how to construct the degenerations case by case, and give explicit definitions of the moduli spaces $\mathcal{X}_J$, the quiver $Q_J$, and the map $\iota_J$. For simplicity, we write $\iota$ omitting $J$ in each subsection. Our strategy is to reduce the number of the cluster $y$-variables keeping the Poisson structure. The automorphism of $\mathcal{X}_J$ is discussed by use of automorphism of the quiver $Q_J$.

\subsection[$\mathcal{P}_{(2,1,1)}$]{$\boldsymbol{\mathcal{P}_{(2,1,1)}}$}

To find the moduli space for a confluence of two punctures, we shall eliminate $y_6$ in $W$~\eqref{W_4puncture}. One sees that the Poisson algebra preserves when we set $y_6 = \frac{\kappa}{y_2 y_4}$. To get a degenerated moduli space we take a limit $\kappa\to \infty$.\footnote{There exist other choices such as $y_6=\kappa y_5$. In $\kappa\to \infty$ we get the same monodromy space with the above by change of variables.} Then dominating terms in~$p_{ab}$ and $p_i$ are written in terms of the cluster $y$-variables $(y_1,\dots,y_5)$. We have
\begin{gather}
 W_{(2,1,1)}(p_{12}, p_{31}, p_{23}; p_1, p_2, p_3) = p_{12} p_{23} p_{31}\nonumber \\
\qquad{}- \big[ {p_{12}}^2 + {p_{31}}^2 + ( p_1 p_2 + p_3 ) p_{12} + ( p_1 p_3 + p_2 ) p_{31} + p_1 p_{23} + p_1 p_2 p_3+ {p_1}^2 + 1 \big], \label{W_211}
\end{gather}
and the realization in terms of the $y$-variables is
\begin{gather}
 \iota(p_1) = \sqrt{\frac{y_1 y_3}{y_2 y_4}}, \qquad \iota(p_2) = \sqrt{y_2 y_3 y_5} + \frac{1}{\sqrt{y_2 y_3 y_5}},\nonumber \\
 \iota(p_3) = \sqrt{y_1 y_4 y_5} +\frac{1}{\sqrt{y_1 y_4 y_5}}, \qquad \iota(p_{12}) = \sqrt{\frac{ y_1 y_5}{y_4}} + \sqrt{\frac{y_1}{y_4 y_5}},\nonumber \\
 \iota(p_{31}) = \sqrt{\frac{y_3 y_5}{y_2}} + \sqrt{\frac{y_5}{y_2 y_3}} +\frac{1}{y_4} \left( \sqrt{\frac{y_3 y_5}{y_2}} +\sqrt{\frac{y_5}{y_2 y_3}} + \frac{1}{\sqrt{y_2 y_3 y_5}} \right),\nonumber \\
 \iota(p_{23}) = \sqrt{y_1 y_2 y_3 y_4} +\sqrt{\frac{y_2 y_3 y_4}{y_1}} +\sqrt{\frac{y_1 y_3 y_4}{y_2}}\nonumber \\
\hphantom{\iota(p_{23}) =}{} + \sqrt{\frac{y_3 y_4}{y_1 y_2}} +\sqrt{\frac{y_4}{y_1 y_2 y_3}} + \sqrt{\frac{y_3}{y_1 y_2 y_4}} + \frac{1}{\sqrt{y_1 y_2 y_3 y_4}} . \label{py_211}
\end{gather}
The polynomial~\eqref{W_211} coincides with one for $P_{\mathrm{V}}$ in~\cite{vdPutSaito09a}.

\subsubsection*{Automorphism}
The automorphism of the quiver $Q_{(2,1,1)}$ is given in terms of the cluster mutations as
\begin{gather} \label{G_211}
 \mathsf{G} = \sigma_{1,2} \sigma_{1,4} \sigma_{2,3} \mu_1 \mu_2 ,
\end{gather}
which is explicitly written as
\begin{gather*}
 \mathsf{G} \colon \
 \begin{pmatrix}
 y_1 \\ y_2 \\ y_3 \\ y_4 \\ y_5
 \end{pmatrix}
 \mapsto
 \begin{pmatrix}
 y_3 (1+y_1 ) ( 1+ y_2 ) \\
 y_4(1+y_1) ( 1+ y_2 ) \\
 {y_2}^{-1} \\
 {y_1}^{-1} \\
 y_5 \big(1+{y_1}^{-1} \big)^{-1} \big( 1+ {y_2}^{-1} \big)^{-1}
 \end{pmatrix} .
\end{gather*}
This induces an action on $\mathcal{X}_{(2,1,1)}$ as
\begin{gather*}
 \mathsf{G}\colon \
 \begin{pmatrix}
 p_1 \\ p_2 \\ p_3
 \end{pmatrix}
 \mapsto
 \begin{pmatrix}
 p_1 \\ p_3 \\ p_2
 \end{pmatrix},
 \qquad
 \begin{pmatrix}
 p_{12} \\ p_{31} \\ p_{23}
 \end{pmatrix}
 \mapsto
 \begin{pmatrix}
 p_{12} p_{23} - p_{31} - p_1 p_3 - p_2 \\
 p_{12} \\
 p_{23}
 \end{pmatrix} .
\end{gather*}

We do have other automorphisms of $Q_{(2,1,1)}$; $\sigma_{2,3} \mu_2 \mu_3 \mu_5$ induces a trivial action on $\mathcal{X}_{(2,1,1)}$, and that $\sigma_{1,4} \sigma_{2,3} \mu_3 \mu_4$ coincides with ${\mathsf{G}}^{-1}$.

\subsection[$\mathcal{P}_{(3,1)}$]{$\boldsymbol{\mathcal{P}_{(3,1)}}$}
In $\mathcal{P}_{(2,1,1)}$, two punctures have been merged by eliminating~$y_6$. We shall further merge one more puncture by setting $ y_1 = \frac{\kappa}{y_4 y_5}$ in~\eqref{py_211} and taking a limit $\kappa \to \infty$. We have
\begin{gather}
 W_{(3,1)}(p_{12}, p_{31}, p_{23}; p_1, p_2)\nonumber \\
\qquad{} = p_{12} p_{23} p_{31} - \big[ {p_{12}}^2 + ( p_1 p_2+1 ) p_{12} +p_1 p_{31} + p_1 p_{23} + p_1 ( p_1 + p_2 ) \big], \label{W_31}
\end{gather}
and an embedding is defined by picking up the dominating term,
\begin{gather}
 \iota(p_1) = \sqrt{\frac{y_3}{y_2 y_5}} {y_4}^{-1},\qquad \iota(p_2) = \sqrt{y_2 y_3 y_5} +\frac{1}{\sqrt{y_2 y_3 y_5}} ,\qquad
 \iota(p_{12}) = {y_4}^{-1} \big( 1 + {y_5}^{-1} \big) ,\nonumber \\
 \iota(p_{31}) = \sqrt{\frac{y_5}{y_2 y_3}} + \sqrt{\frac{y_3 y_5}{y_2}} +\frac{1}{y_4} \left( \sqrt{\frac{y_3 y_5}{y_2}} + \sqrt{\frac{y_5}{y_2 y_3}} + \frac{1}{\sqrt{y_2 y_3 y_5}} \right) , \nonumber\\
 \iota(p_{23}) = \sqrt{\frac{y_3}{y_2 y_5}} + \sqrt{\frac{y_2 y_3}{y_5}} . \label{py_31}
\end{gather}
The moduli space~\eqref{W_31} coincides with $P_{\mathrm{IV}}$
in~\cite{vdPutSaito09a},
\begin{gather*}
 W^\prime_{(3,1)}(p_{12}, p_{31}, p_{23}; p_1, p_2) \\
\qquad{} = p_{12} p_{23} p_{31} - \big[ {p_{12}}^2 + \big( {p_1}^{-1} p_2+ {p_1}^{-2} \big) p_{12} +{p_1}^{-2} p_{31} + {p_1}^{-2} p_{23} + {p_1}^{-2}+{p_1}^{-3} p_2 \big] ,
\end{gather*}
when we re-parametrize $p_{12}$, $p_{31}$, $p_{23}$ with ${p_1}^2 p_{12}$, $p_1 p_{31}$, $p_1 p_{23}$, respectively.

\subsubsection*{Automorphism}
The automorphism of $Q_{(3,1)}$ is given by
\begin{gather} \label{G_31}
 \mathsf{G} = \sigma_{2,5} \sigma_{3,4} \mu_3 \mu_4 ,
\end{gather}
which is explicitly written as
\begin{gather*}
 \mathsf{G}\colon \
 \begin{pmatrix}
 y_2 \\ y_3 \\ y_4 \\ y_5
 \end{pmatrix}
 \mapsto
 \begin{pmatrix}
 y_5 ( 1+ y_3 ) ( 1+ y_4 ) \\
 {y_4}^{-1} \\
 {y_3}^{-1} \\
 y_2 \big( 1+ {y_3}^{-1} \big)^{-1} \big( 1+ {y_4}^{-1} \big)^{-1}
 \end{pmatrix} .
\end{gather*}
Then the induced action on $\mathcal{X}_{(3,1)}$ is
\begin{gather*}
 \mathsf{G}\colon \
 \begin{pmatrix}
 p_1 \\
 p_2
 \end{pmatrix}
 \mapsto
 \begin{pmatrix}
 p_1 \\
 p_2
 \end{pmatrix} , \qquad
 \begin{pmatrix}
 p_{12} \\ p_{31} \\ p_{23}
 \end{pmatrix}
 \mapsto
 \begin{pmatrix}
 p_{23} p_{31} - p_{12} - p_1 p_2 - 1 \\
 p_{23} \\
 p_{31}
 \end{pmatrix} .
\end{gather*}

\subsection[$\mathcal{P}_{(2,2)}$]{$\boldsymbol{\mathcal{P}_{(2,2)}}$}

To merge remaining two punctures in $\mathcal{P}_{(2,1,1)}$~\eqref{py_211} together, we put $ y_5 = \frac{\kappa}{y_1 y_4}$ and take a~limit $\kappa\to \infty$. Extracting dominating terms in the embedding~$\iota$, all are written in terms of $(y_1, y_2, y_3, y_4)$. We have
\begin{gather}
 W_{(2,2)}(p_{12}, p_{31}, p_{23}; p_1, p_2)\nonumber \\
 \qquad {}= p_{12} p_{23} p_{31} - \big[ {p_{12}}^2 + {p_{31}}^2 + ( p_1 p_2+ 1 ) p_{12} + ( p_1 + p_2 ) p_{31} + p_1 p_2 \big], \label{W_22}
\end{gather}
 and $p$'s are realized as
\begin{gather}
 \iota(p_1) = \sqrt{\frac{y_1 y_3}{y_2 y_4}} ,\qquad \iota(p_2) = \sqrt{\frac{y_2 y_3}{y_1 y_4}},
\qquad \iota(p_{12}) = {y_4}^{-1} ,\nonumber \\
 \iota(p_{31}) = \frac{1}{\sqrt{y_1 y_2 y_3 y_4}} ( 1 + y_3 ) \big( 1 + {y_4}^{-1} \big) ,\nonumber \\
 \iota(p_{23}) = \sqrt{y_1 y_2 y_3 y_4} +\sqrt{\frac{y_2 y_3 y_4}{y_1}} +\sqrt{\frac{y_1 y_3 y_4}{y_2}}\nonumber \\
\hphantom{\iota(p_{23}) =}{} + \sqrt{\frac{y_3 y_4}{y_1 y_2}} +\sqrt{\frac{y_4}{y_1 y_2 y_3}} + \sqrt{\frac{y_3}{y_1 y_2 y_4}} + \frac{1}{\sqrt{y_1 y_2 y_3 y_4}} . \label{py_22}
\end{gather}
The polynomial~\eqref{W_22} coincides with one for $P_{\mathrm{III}}^{D_6}$ in~\cite{vdPutSaito09a}.

\subsubsection*{Automorphism}
The automorphism of the quiver $Q_{(2,2)}$ is given by
\begin{gather} \label{G_22}
 \mathsf{G} = \sigma_{1,3} \sigma_{2,4} \mu_1 \mu_2 ,
\end{gather}
which is explicitly written as
\begin{gather*}
 \mathsf{G}\colon \
 \begin{pmatrix}
 y_1 \\ y_2 \\ y_3 \\ y_4
 \end{pmatrix}
 \mapsto
 \begin{pmatrix}
 y_3( 1+y_1 ) (1+y_2 ) \\
 y_4 ( 1+y_1)(1+y_2 ) \\
 {y_1}^{-1} \\
 {y_2}^{-1}
 \end{pmatrix} .
\end{gather*}
We can check that this induces an action on $\mathcal{X}_{(2,2)}$ as
\begin{gather*}
 \mathsf{G}\colon \
 \begin{pmatrix}
 p_1 \\
 p_2
 \end{pmatrix}
 \mapsto
 \begin{pmatrix}
 p_2 \\
 {p_1}^{-1}
 \end{pmatrix},
 \qquad
 \begin{pmatrix}
 p_{12} \\
 p_{31} \\
 p_{23}
 \end{pmatrix}
 \mapsto
 \begin{pmatrix}
 {p_1}^{-1}( p_{12} p_{23} - p_{31} - p_1 - p_2 ) \\
 {p_1}^{-1} p_{12} \\
 p_{23}
 \end{pmatrix} .
\end{gather*}

\subsection[$\mathcal{P}_{(2,1,1)}^{\text{deg}}$]{$\boldsymbol{\mathcal{P}_{(2,1,1)}^{\text{deg}}}$}

As the moduli space of $\mathcal{P}_{(2,1,1)}$~\eqref{W_211} coincides with one for $P_{\mathrm{V}}$ in~\cite{vdPutSaito09a}, we shall pay attention to a degeneration~$p_1=0$ in~\eqref{py_211}. For this purpose, we renormalize cluster variables $y_2$ and $y_3$ by~$\frac{y_2}{\varepsilon}$ and $\varepsilon y_3$ respectively, and take a limit $\varepsilon\to 0$. Then we introduce a new cluster variable by $y_{2^\prime}=y_2 y_3$. As a result, we have a quiver as in Fig.~\ref{fig:degenerate}, and the moduli space is
\begin{gather} \label{W_211deg}
 W_{(2,1,1)^{\text{deg}}}( p_{12}, p_{31}, p_{23} ; p_2, p_3) = p_{12} p_{23} p_{31} - \big[ {p_{12}}^2 + {p_{31}}^2 + p_3 p_{12} + p_2 p_{31} + 1 \big] .
\end{gather}
We have
\begin{gather}
 \iota(p_2) = \sqrt{ y_{2^\prime} y_5} + \frac{1}{\sqrt{y_{2^\prime} y_5}} , \qquad \iota(p_3) = \sqrt{y_1 y_4 y_5} + \frac{1}{\sqrt{y_1 y_4 y_5}} ,\qquad
 \iota(p_{12}) = \sqrt{\frac{y_1}{y_4 y_5}} ( 1 + y_5) , \nonumber\\
 \iota(p_{31}) = \sqrt{\frac{y_5}{y_{2^\prime}}} + {y_4}^{-1} \frac{1}{\sqrt{y_{2^\prime} y_5}} ( 1+y_5 ),\nonumber \\
 \iota(p_{23}) = \sqrt{y_1 y_{2^\prime} y_4} +\sqrt{\frac{y_{2^\prime} y_4}{y_1}} +\sqrt{\frac{y_4}{y_1 y_{2^\prime}}} +\frac{1}{\sqrt{y_1 y_{2^\prime} y_4 }} .\label{py_211deg}
\end{gather}
The polynomial~\eqref{W_211deg} is indeed the case for $P_{\mathrm{V}}^{\text{deg}}$ in~\cite{vdPutSaito09a}.

\subsubsection*{Automorphism}
The automorphism of the quiver $Q_{(2,1,1)^{\text{deg}}}$ is given by
\begin{gather} \label{G_211deg}
 \mathsf{G} = \sigma_{1,2^\prime} \sigma_{1,4} \mu_1 ,
\end{gather}
which is explicitly written as
\begin{gather*}
 \mathsf{G} \colon \
 \begin{pmatrix}
 y_1 \\ y_{2^\prime} \\ y_4 \\ y_5
 \end{pmatrix}
 \mapsto
 \begin{pmatrix}
 y_{2^\prime} ( 1 + y_1 ) \\
 y_4 ( 1 + y_1 ) \\
 {y_1}^{-1} \\
 y_5 \big( 1+{y_{1}}^{-1} \big)^{-1}
 \end{pmatrix} .
\end{gather*}
We see that this induces an action on $\mathcal{X}_{(2,1,1)^{\text{deg}}}$,
\begin{gather*}
 \mathsf{G} \colon \
 \begin{pmatrix}
 p_2 \\ p_3
 \end{pmatrix}
 \mapsto
 \begin{pmatrix}
 p_3 \\ p_2
 \end{pmatrix} ,
 \qquad
 \begin{pmatrix}
 p_{12} \\ p_{31} \\ p_{23}
 \end{pmatrix}
 \mapsto
 \begin{pmatrix}
 p_{12} p_{23} - p_{31} - p_2 \\
 p_{12} \\
 p_{23}
 \end{pmatrix} .
\end{gather*}

\subsection[$\mathcal{P}_{(4)A}$]{$\boldsymbol{\mathcal{P}_{(4)A}}$}
In $\mathcal{P}_{(3,1)}$~\eqref{py_31}, we set $ y_3 = \frac{\kappa}{y_2 y_5}$ and take a limit $\kappa \to \infty$. We get
\begin{gather} \label{W_4A}
 W_{(4)A}(p_{12}, p_{31}, p_{23}; p_1) = p_{12} p_{23} p_{31} - [ p_1 ( p_{12} + p_{23} + p_{31} ) + p_1 ( p_1 + 1 ) ],
\end{gather}
and we have a cyclic presentation
\begin{gather}
 \iota(p_1) = \frac{1}{y_2 y_4 y_5},\qquad \iota(p_{12}) ={y_4}^{-1} \big( 1+{y_5}^{-1} \big) , \qquad \iota( p_{31}) = {y_2}^{-1} \big( 1 +{y_4}^{-1} \big) , \nonumber\\
 \iota(p_{23}) = {y_5}^{-1} \big( 1 + {y_2}^{-1} \big) . \label{py_4A}
\end{gather}
When we replace $p_{12}$ by $ p_1 p_{12}$, we recover the same result with one for $P_{\mathrm{II}}^{\text{JM}}$ in~\cite{vdPutSaito09a}
\begin{gather*}
 {W}^\prime_{(4)A}( p_{12}, p_{31}, p_{23} ; p_1 ) = p_{12} p_{23} p_{31} - [ p_1 p_{12}+p_{31} + p_{23}+ p_1 +1 ] .
\end{gather*}

\subsection[$\mathcal{P}_{(4)B}$]{$\boldsymbol{\mathcal{P}_{(4)B}}$}
In $\mathcal{P}_{(3,1)}$~\eqref{py_31}, we set $ y_5 = \frac{\kappa}{y_2 y_3}$ and take a limit $\kappa\to \infty$. We have
\begin{gather} \label{W_4B}
 W_{(4)B}(p_{12}, p_{31}, p_{23}; p_1) = p_{12} p_{23} p_{31} - \big[ {p_{12}}^2 + ( p_1+1 ) p_{12} + p_1 p_{31} +p_1 \big],
\end{gather}
and $p$'s are written in terms of $(y_2, y_3, y_4)$ as
\begin{gather*}
 \iota(p_1) = y_3 {y_4}^{-1},\qquad \iota(p_{12}) = {y_4}^{-1} , \qquad \iota(p_{31}) = {y_2}^{-1} \big(1+{y_3}^{-1}\big) \big(1+{y_4}^{-1}\big) , \\
 \iota(p_{23}) = ( 1+ y_2 ) y_3 ,
\end{gather*}
where we have extracted the dominating terms in the limit.

The same moduli space can be given from $\mathcal{P}_{(2,2)}$. In~\eqref{py_22} we set $y_1=\kappa y_2$ and take a limit $\kappa\to\infty$. We have
\begin{gather} \label{W_4B_2}
 W^\prime_{(4)B}(p_{12}, p_{31}, p_{23} ; p_1) = p_{12} p_{23} p_{31} - \big[ {p_{12}}^2 + \big( {p_1}^2 +1 \big) p_{12} + p_1 p_{31} + {p_1}^2 \big] ,
\end{gather}
and an embedding is defined by
\begin{gather*}
\iota(p_1) = \sqrt{\frac{y_3}{y_4}} ,\qquad \iota(p_{12}) = {y_4}^{-1} ,\qquad \iota(p_{31}) = \frac{1}{y_2 \sqrt{y_3 y_4}} (1+y_3 ) \big( 1+ {y_4}^{-1} \big) , \\
 \iota(p_{23}) = (1+y_2) \sqrt{y_3 y_4} .
\end{gather*}
We see that $W^\prime_{(4)B}$ is equivalent with $W_{(4)B}$~\eqref{W_4B} by replacing $p_{31}$ and $p_{23}$ by $p_1 p_{31}$ and $\frac{p_{23}}{p_1}$ respectively.

Also we may put $ y_3 = \kappa y_4$ in $\mathcal{P}_{(2,2)}$~\eqref{py_22} and take a limit $\kappa\to\infty$. In this process, the quiver is not $Q_{(4)B}$ in Fig.~\ref{fig:degenerate}, but a mutated one, $\mu_2(Q_{(4)B})$. The moduli space is now
\begin{gather} \label{W_4B_3}
 W_{(4)B}^{\prime\prime}(p_{12}, p_{31}, p_{23}; p_1) = p_{12} p_{23} p_{31} - \big[ p_{12} + {p_{31}}^2 +\big( p_1+{p_1}^{-1} \big) p_{31} + 1 \big],
\end{gather}
and
\begin{gather*}
 \iota(p_1) = \sqrt{\frac{y_1}{y_2}} , \qquad \iota(p_{12}) = {y_4}^{-1},\qquad \iota(p_{31}) = \frac{1}{\sqrt{y_1 y_2}} \big( 1+{y_4}^{-1} \big) ,\\
 \iota(p_{23}) = \frac{1}{\sqrt{y_1 y_2}} + \frac{1}{\sqrt{y_1 y_2}} ( 1+y_1 ) ( 1+y_2 ) y_4 .
\end{gather*}
It is seen that, by replacing $p_{12}$ and $p_{31}$ with $p_1 p_{12}$ and $p_1 p_{31}$ respectively in~\eqref{W_4B_3}, $W^{\prime\prime}_{(4)B}$ is equivalent to $W^\prime_{(4)B}$~\eqref{W_4B_2}.

\subsection[$\mathcal{P}_{(2,2)}^{\text{deg}}$]{$\boldsymbol{\mathcal{P}_{(2,2)}^{\text{deg}}}$}
Having seen that $\mathcal{P}_{(2,2)}$~\eqref{py_22} is the moduli space for $P_{\mathrm{III}}^{D_6}$ studied in~\cite{vdPutSaito09a}, we shall pay attention to a degenerated case, $p_1=0$, in $\mathcal{P}_{(2,2)}$. To realize this case, we renormalize $y_2$ and $y_3$ by $\frac{y_2}{\varepsilon}$ and $\varepsilon y_3$ respectively, and take a limit $\varepsilon\to 0$. Then by introducing a new variable $y_{2^\prime} = y_2 y_3$, we obtain
\begin{gather*}
 W_{(2,2)^{\text{deg}}}(p_{12}, p_{31}, p_{23}; p_2) = p_{12} p_{23} p_{31} - \big[ {p_{12}}^2 + {p_{31}}^2 + p_{12} + p_2 p_{31} \big] ,
\end{gather*}
and the realization is given by
\begin{gather}
 \iota(p_2) = \sqrt{\frac{y_{2^\prime}}{y_1 y_4}} ,\qquad \iota(p_{12}) = {y_4}^{-1}, \qquad
 \iota(p_{31}) = \frac{1}{\sqrt{y_1 y_{2^\prime} y_4}} \big(1+{y_4}^{-1}\big),\nonumber \\
 \iota(p_{23}) = \big( 1 + {y_1}^{-1} \big) \sqrt{y_1 y_{2^\prime} y_4} + \frac{1}{\sqrt{y_1 y_{2^\prime} y_4} }(1+y_4) . \label{py_22deg}
\end{gather}
This is equivalent to the moduli space for $P_{\mathrm{III}}^{D_7}$ in~\cite{vdPutSaito09a}.

It is noted that we obtain the same result from $\mathcal{P}_{(2,1,1)}^{\text{deg}}$~\eqref{py_211deg} by setting $y_5=\frac{\kappa}{y_1 y_4}$ and taking a limit $\kappa\to\infty$.

\subsubsection*{Automorphism}
The automorphism of the quiver $Q_{(2,2)^{\text{deg}}}$ is
\begin{gather*}
 \mathsf{G} = \sigma_{1,4} \sigma_{1,2^\prime} \mu_1 ,
\end{gather*}
which is explicitly written as
\begin{gather*}
 \mathsf{G}\colon \
 \begin{pmatrix}
 y_1 \\ y_{2^\prime} \\ y_4
 \end{pmatrix}
 \mapsto
 \begin{pmatrix}
 y_{2^\prime} ( 1 + y_1) \\
 y_4 ( 1 + y_1) \\
 {y_1}^{-1}
 \end{pmatrix} .
\end{gather*}
This induces
\begin{gather*}
 \mathsf{G}\colon \
 p_2 \mapsto {p_2}^{-1} , \qquad
 \begin{pmatrix}
 p_{12} \\ p_{31} \\ p_{23}
 \end{pmatrix}
 \mapsto
 \begin{pmatrix}
 {p_2}^{-1} p_{12} p_{23} - {p_2}^{-1} p_{31} - 1 \\
 {p_2}^{-1} p_{12} \\
 p_{23}
 \end{pmatrix} .
\end{gather*}

\subsection[$\mathcal{P}_{(2,2)}^{\text{deg}^2}$]{$\boldsymbol{\mathcal{P}_{(2,2)}^{\text{deg}^2}}$}
We have interests in a further degenerated case, $p_2=0$, in the previous case $\mathcal{P}_{(2,2)}^{\text{deg}}$~\eqref{py_22deg}. We replace $y_1$ and $y_{2^\prime}$ by $\frac{y_1}{\varepsilon}$ and $\varepsilon y_{2^\prime}$ respectively, and set $\varepsilon=0$. By introducing a new variable $y_{2^{\prime\prime}}=y_1 y_{2^\prime}$, we have the Kronecker quiver as $Q_{(2,2)}^{\text{deg}^2}$. The moduli space is
\begin{gather*}
 W_{(2,2)^{\text{deg}^2}}( p_{12}, p_{31}, p_{23} ) = p_{12} p_{23} p_{31} - \big[ {p_{12}}^2 + {p_{31}}^2 +p_{12} \big] ,
\end{gather*}
and the embedding is
\begin{gather*}
 \iota(p_{12}) = {y_4}^{-1}, \qquad \iota(p_{31}) = \frac{1}{\sqrt{y_{2^{\prime\prime}} y_4}} \big( 1 + {y_4}^{-1} \big) , \\
\iota(p_{23}) = \sqrt{y_{2^{\prime\prime}} y_4} + \frac{1}{\sqrt{y_{2^{\prime\prime}} y_4}} ( 1 + y_4 ) .
\end{gather*}
The polynomial $W_{(2,2)}^{\text{deg}^2}$ coincides with $P_{\mathrm{III}}^{D_8}$ in~\cite{vdPutSaito09a}.

\subsection[$\mathcal{P}_{(3,1)B}$]{$\boldsymbol{\mathcal{P}_{(3,1)B}}$}
We set $y_4=\frac{\kappa}{y_2 y_5}$ in $\mathcal{P}_{(3,1)}$~\eqref{py_31} and take a limit $\kappa\to\infty$. Then we have
\begin{gather} \label{W_31B}
 W_{(3,1)B}(p_{12}, p_{31}, p_{23} ; p_1 ) = p_{12} p_{23} p_{31} - \big[ p_{12} + p_1 p_{31} + p_1 p_{23} + {p_1}^2+ 1 \big],
\end{gather}
and $p$'s are given in terms of $(y_2, y_3, y_5)$ as
\begin{gather*}
 \iota(p_1) = \sqrt{y_2 y_3 y_5} , \qquad \iota(p_{12}) = y_2 ( 1 + y_5) ,\qquad \iota(p_{31}) = \sqrt{\frac{y_5}{y_2 y_3}} ( 1 + y_3) , \\
\iota(p_{23}) = \sqrt{\frac{y_3}{y_2 y_5}} ( 1 + y_2) .
\end{gather*}
We see, by replacing $p_{31}$ and $p_{23}$ with $\frac{p_{31}}{p_1}$ and $\frac{p_{23}}{p_1}$ respectively in~\eqref{W_31B}, that the moduli space is equivalent to $W_{(4)A}$~\eqref{W_4A}.

From $\mathcal{P}_{(2,1,1)}^{\text{deg}}$~\eqref{W_211deg}, we put $y_{2^\prime} = \kappa y_1 y_4$ and take a limit $\kappa\to\infty$. We obtain
\begin{gather*}
 W^\prime_{(3,1)B}(p_{12}, p_{31}, p_{23}; p_2) = p_{12} p_{23} p_{31} - \big[ {p_{12}}^2 + \big(p_2 + {p_2}^{-1} \big) p_{12} + p_2 p_{31} + 1 \big],
\end{gather*}
and the realization is
\begin{gather*}
 \iota(p_2) = \sqrt{ y_1 y_4 y_5},\qquad \iota(p_{12}) = \sqrt{\frac{y_1}{y_4 y_5}} ( 1+y_5) , \\
\iota(p_{31}) = \sqrt{\frac{y_5}{y_1 y_4}} + \frac{1}{\sqrt{y_1 y_4 y_5}} {y_4}^{-1} ( 1 + y_5 ) , \qquad \iota(p_{23}) = ( 1 + y_1 ) y_4 .
\end{gather*}
We see that the polynomial is essentially equivalent to $W^{\prime\prime}_{(4)B}$~\eqref{W_4B_3} by replacing~$p_{31}$ and~$p_{23}$ with $\frac{p_{31}}{p_2}$ and~$p_2 p_{23}$ respectively. It is also seen that, by replacing $p_{12}$, $p_{31}$, $p_{23}$ with $\frac{p_{12}}{p_2}$, $\frac{1}{{p_2}^2} ( p_{12} p_{31} - p_2)$, $p_2 p_{23}$ respectively, the polynomial $W^\prime_{(3,1)B}$ is essentially equivalent to $W_{(3,1)B}$~\eqref{W_31B}.

\subsection[$\mathcal{P}_{(4)C}$]{$\boldsymbol{\mathcal{P}_{(4)C}}$}
In $\mathcal{P}_{(4)A}$~\eqref{py_4A}, we set $y_4=\frac{\kappa}{y_2 y_5}$ and take a~limit $\kappa \to \infty$. We have
\begin{gather*}
 W_{(4)C}(p_{12}, p_{31}, p_{23}) = p_{12} p_{23} p_{31} - [ p_{31}+ p_{23} + 1 ] ,
\end{gather*}
and
\begin{gather*}
 \iota(p_{12}) = y_2 ( 1 + y_5 ) ,\qquad \iota(p_{31}) = {y_2}^{-1}, \qquad \iota(p_{23}) = \big( 1+{y_2}^{-1} \big) {y_5}^{-1} .
\end{gather*}
This is same with the case of $P_{\mathrm{I}}$ in~\cite{vdPutSaito09a}.

We also have the same moduli space from $\mathcal{P}_{(4)B}$~\eqref{W_4B} by setting $y_4 = \kappa y_3$ and taking a limit $\kappa \to \infty$. From $\mathcal{P}_{(3,1)B}$~\eqref{W_31B}, we may put $y_3=\frac{\kappa}{y_2 y_5}$ and take a limit $\kappa\to\infty$.

\subsection{Remarks}
We have $11$ diagrams as confluences of the 4~punctures on the sphere. Some of them have quivers which are mutation-equivalent. Apparently $Q_{(4)A}$ and $Q_{(3,1)B}$ are same, and they are mutation-equivalent to $Q_{(4)B}$.
Quivers $Q_{(2,1,1)}^{\text{deg}}$ and $Q_{(2,2)}$ are also mutation equivalent. All quivers $Q_J$ are finite mutation type. Classification of finite mutation type quivers were studied in~\cite{FeliShapTuma12a, FomiShapThur08a}, andvshown is that they can be constructed by gluing fundamental quivers, called ``blocks'' of type-I, II, IIIa, IIIb, IV, V, with some exceptions. One sees that our $Q_J$ are the blocks.

As seen in the previous subsections, some of the moduli spaces for $\mathcal{P}_J$ are equivalent. One also finds that the polynomial $W_{(2,2)}$~\eqref{W_22} is essentially equivalent to $W_{(2,1,1)}^{\text{deg}}$~\eqref{W_211deg} when we replace $p_{12}$ and $p_{31}$ by $\sqrt{p_1 p_2} p_{12}$ and $\sqrt{p_1 p_2} p_{31}$ respectively.

\section{Quantization of character varieties: once-punctured torus}\label{sec:q_torus}

As we have seen that the trace map relates the cluster algebra with the character variety, we apply the quantum cluster algebra~\cite{BerenZelev05a,FockGonc09a} to quantize the character variety of the once-punctured torus.

To give a quantization of $\mathcal{X}(\Sigma_{1,1})$, we set $W_{\text{Mar}}^q\big(\hat{x},\hat{y},\hat{z}; \hat{b}\big)$ by
\begin{gather*}
 W_{\text{Mar}}^q\big(\hat{x},\hat{y},\hat{z}; \hat{b}\big) = q^{\frac{1}{2}} \hat{x} \hat{y} \hat{z} - q {\hat{x}}^2 - q^{-1} {\hat{y}}^2 - q {\hat{z}}^2 + q+q^{-1} - \hat{b} .
\end{gather*}
Then we define a quantized moduli space $\mathcal{X}^q(\Sigma_{1,1})$ as generated by $\hat{x}$, $\hat{y}$, and $\hat{z}$, with a quotient by a quantized affine surface
\begin{gather*}
 W_{\text{Mar}}^q\big(\hat{x}, \hat{y}, \hat{z}; \hat{b}\big)=0 ,
\end{gather*}
and the quantum algebras
 \begin{gather*}
 q^{\frac{1}{2}} \hat{x} \hat{y} - q^{-\frac{1}{2}} \hat{y} \hat{x} = \big( q-q^{-1} \big) \hat{z} , \\
 q^{\frac{1}{2}} \hat{y} \hat{z} - q^{-\frac{1}{2}} \hat{z} \hat{y} = \big( q-q^{-1} \big) \hat{x} , \\
 q^{\frac{1}{2}} \hat{z} \hat{x} - q^{-\frac{1}{2}} \hat{x} \hat{z} = \big( q-q^{-1} \big) \hat{y} .
\end{gather*}
Here $\hat{b}$ is a center
\begin{gather*}
 \hat{x} \hat{b} = \hat{b} \hat{x} , \qquad \hat{y} \hat{b} = \hat{b} \hat{y} , \qquad \hat{z} \hat{b} = \hat{b} \hat{z} .
\end{gather*}
This algebra was identified with the Kauffman bracket skein module on the once-punctured torus~\cite{BulloPrzyt99a}. It was also realized in~\cite{Oblom04a} by use of the (spherical) double affine Hecke algebra of type $A_1$, and we have a representation in terms of the $A_1$ Macdonald polynomial, i.e., the $q$-ultraspherical polynomial.

The map in Theorem~\ref{thm:torus_phi} is naturally quantized as follows. Here we have the quantum cluster algebra $\mathcal{A}^q(Q_{\text{Mar}})$ generated by $Y_1$, $Y_2$, and $Y_3$ satisfying~\eqref{quantum_Y} with the exchange matrix~\eqref{B_torus}.

\begin{Theorem} \label{thm:q_torus} We have an embedding of algebra
 \begin{gather*}
 \iota\colon \ \mathcal{X}^q(\Sigma_{1,1}) \to \mathcal{A}^q(Q_{\text{\rm Mar}})
 \end{gather*}
 defined by
 \begin{gather}
\iota(\hat{b}) = q^{-2} Y_1 Y_2 Y_3 + q^2 ( Y_1 Y_2 Y_3 )^{-1} ,\nonumber \\
 \iota(\hat{x}) = q^{-\frac{1}{2}} {Y_3}^{\frac{1}{2}} {Y_1}^{\frac{1}{2}} + q^{\frac{1}{2}} {Y_3}^{-\frac{1}{2}} {Y_1}^{\frac{1}{2}} + q^{-\frac{1}{2}} {Y_3}^{-\frac{1}{2}} {Y_1}^{-\frac{1}{2}} ,\nonumber \\
 \iota(\hat{y}) = q^{-\frac{1}{2}} {Y_2}^{\frac{1}{2}} {Y_3}^{\frac{1}{2}} + q^{\frac{1}{2}} {Y_2}^{-\frac{1}{2}} {Y_3}^{\frac{1}{2}} + q^{-\frac{1}{2}} {Y_2}^{-\frac{1}{2}} {Y_3}^{-\frac{1}{2}} ,\nonumber \\
 \iota(\hat{z}) = q^{-\frac{1}{2}} {Y_1}^{\frac{1}{2}} {Y_2}^{\frac{1}{2}} + q^{\frac{1}{2}} {Y_1}^{-\frac{1}{2}} {Y_2}^{\frac{1}{2}} + q^{-\frac{1}{2}} {Y_1}^{-\frac{1}{2}} {Y_2}^{-\frac{1}{2}} .\label{quantum_xyz}
 \end{gather}
\end{Theorem}

This can be checked by the $q$-commuting relations~\eqref{quantum_Y} with~\eqref{B_torus}. As in the classical case~\eqref{xyz_y}, we follow the quantum trace for the Kauffman bracket skein module on $\Sigma_{1,1}$ to construct the map~$\iota$. Here the quantum $Y$-variables are defined in terms of the quantum triangle algebra $\mathcal{T}^q_\Delta$ generated by $Z_1$, $Z_2$, and $Z_3$, satisfying
\begin{gather*}
 {Z}_j {Z}_{j+1} = q^2 {Z}_{j+1} {Z}_j ,
\end{gather*}
for periodic $j\in \{ 1, 2, 3 \}$. The Boltzmann weight~\eqref{gamma_arc} for each arc is quantized to be
\begin{gather*}
 \gamma_\Delta(\epsilon, \epsilon^\prime) =
 \begin{cases}
 0 , & \text{if $(\epsilon, \epsilon^\prime)=(-1, +1)$}, \\
 q^{-\frac{1}{4} \epsilon \epsilon^\prime} {Z_{ j}}^{\frac{1}{2}\epsilon} {Z_{ j+1}}^{\frac{1}{2}\epsilon^\prime}, & \text{otherwise}.
 \end{cases}
\end{gather*}
In our cases of the punctured torus, the triangulation is simple, and all loops, $\mathbb{x}$, $\mathbb{y}$, and $\mathbb{z}$, are in good positions. The map~\eqref{quantum_xyz} follows by taking a trace of the above weight for each loop, and it should be compared to expectation values of a supersymmetric line defect~\cite{GaioMoorNeit13a}.

The automorphism of $\mathcal{X}^q(\Sigma_{1,1})$ is given from quantum mutations. The quantization of the mutations~\eqref{RL_torus}, $ \mathsf{R} = \sigma_{1,3} \mu^q_1$ and $ \mathsf{L} = \sigma_{2,3} \mu^q_2 $, on~$\mathcal{A}^q(Q_{\text{oct}})$ is read as
\begin{gather*}
 \mathsf{R}\colon \ \begin{pmatrix}
 Y_1 \\ Y_2 \\Y_3
 \end{pmatrix}
 \mapsto
 \begin{pmatrix}
 Y_3 \big( 1 + q {Y_1}^{-1}\big)^{-1} \big( 1+q^3 {Y_1}^{-1} \big)^{-1} \\
 Y_2 ( 1+q Y_1) \big( 1+q^3 Y_1 \big) \\
 {Y_1}^{-1}
 \end{pmatrix}, \\
 \mathsf{L}\colon \
 \begin{pmatrix}
 Y_1 \\ Y_2 \\Y_3
 \end{pmatrix}
 \mapsto
 \begin{pmatrix}
 Y_1 \big( 1 + q {Y_2}^{-1}\big)^{-1} \big( 1+q^3 {Y_2}^{-1} \big)^{-1} \\
 Y_3 (1+q Y_2) \big( 1+q^3 Y_2 \big) \\
 {Y_2}^{-1}
 \end{pmatrix}.
\end{gather*}
One sees that the braid relation is fulfilled, $\mathsf{R} \circ \mathsf{L}^{-1} \circ \mathsf{R} =\mathsf{L}^{-1} \circ \mathsf{R} \circ \mathsf{L}^{-1}$. These actions induce the $\mathrm{PSL}(2;\mathbb{Z})$ actions on $\mathcal{X}^q(\Sigma_{1,1})$ as follows.
\begin{Proposition} We have
 \begin{gather*}
 \mathsf{R}\colon \ \begin{pmatrix}
 \hat{x} \\
 \hat{y} \\
 \hat{z}
 \end{pmatrix}
 \mapsto
 \begin{pmatrix}
 \hat{x} \\
 \hat{z} \\
 q^{\frac{1}{2}} \hat{z} \hat{x} - q \hat{y}
 \end{pmatrix} , \qquad
 \hat{b} \mapsto \hat{b}, \\
 \mathsf{L}\colon \ \begin{pmatrix}
 \hat{x} \\
 \hat{y} \\
 \hat{z}
 \end{pmatrix}
 \mapsto
 \begin{pmatrix}
 \hat{z} \\
 \hat{y} \\
 q^{\frac{1}{2}} \hat{y} \hat{z} - q \hat{x}
 \end{pmatrix}, \qquad
 \hat{b} \mapsto \hat{b}.
 \end{gather*}
\end{Proposition}

\section{Quantization of character varieties: 4-punctured sphere}\label{sec:q_sphere}

We shall formulate a quantization of the character variety of the $4$-punctured sphere. As we already have a cluster algebraic formulation (not a ``general'' cluster algebra) in Theorem~\ref{thm:map_sphere}, we can employ
the quantum cluster algebra $\mathcal{A}^q(Q_{\text{oct}})$ generated by the $q$-commuting variables $Y_1,\dots,Y_6$ satisfying~\eqref{quantum_Y} with the exchange matrix~\eqref{B_sphere}.

We set a quantized affine surface by
\begin{gather} \label{qW_is_zero}
 W^q \big( \hat{p}_{12}, \hat{p}_{31}, \hat{p}_{23}; \hat{p}_1, \hat{p}_2, \hat{p}_3, \hat{p}_4 \big) = 0,
\end{gather}
where
\begin{gather}
 W^q \big( \hat{p}_{12}, \hat{p}_{31}, \hat{p}_{23}; \hat{p}_1, \hat{p}_2, \hat{p}_3, \hat{p}_4 \big) = q^{-1} \hat{p}_{12} \hat{p}_{23} \hat{p}_{31} - \bigl[ q^{-2} {\hat{p}_{12}}^{2} + q^{-2} {\hat{p}_{31}}^{2}
 + q^2 {\hat{p}_{23}}^{2} \nonumber\\
 \qquad {} +q^{-1} \big( \hat{p}_1 \hat{p}_2 + \hat{p}_3 \hat{p}_4 \big) \hat{p}_{12} +q^{-1} \big( \hat{p}_1 \hat{p}_3 + \hat{p}_2 \hat{p}_4 \big) \hat{p}_{31} +q \big( \hat{p}_1 \hat{p}_4 + \hat{p}_2 \hat{p}_3 \big)
 \hat{p}_{23}\nonumber \\
 \qquad{} + {\hat{p}_1}^{2} + {\hat{p}_2}^{2}+ {\hat{p}_3}^{2} + {\hat{p}_4}^{2} + \hat{p}_1 \hat{p}_2 \hat{p}_3 \hat{p}_4 -\big( q+q^{-1} \big)^2 \bigr]. \label{center_x_b}
\end{gather}
Then we define a quantized character variety $\mathcal{X}^q(\Sigma_{0,4})$ as algebra generated by $\hat{p}_{12}$, $\hat{p}_{31}$, and $\hat{p}_{23}$ with a~quotient by $W^q$ and the commutation relations
\begin{gather}
 q \hat{p}_{12} \hat{p}_{31} - q^{-1} \hat{p}_{31} \hat{p}_{12} = \big(q^2-q^{-2}\big) \hat{p}_{23} + \big(q-q^{-1}\big) \big( \hat{p}_1 \hat{p}_4 + \hat{p}_2 \hat{p}_3 \big) ,\nonumber \\
q \hat{p}_{31} \hat{p}_{23} - q^{-1} \hat{p}_{23} \hat{p}_{31} = \big(q^2-q^{-2}\big) \hat{p}_{12} + \big(q-q^{-1}\big) \big( \hat{p}_1 \hat{p}_2 + \hat{p}_3 \hat{p}_4 \big) ,\nonumber \\
 q \hat{p}_{23} \hat{p}_{12} - q^{-1} \hat{p}_{12} \hat{p}_{23} = \big(q^2-q^{-2}\big) \hat{p}_{31} + \big(q-q^{-1}\big) \big( \hat{p}_1 \hat{p}_3 + \hat{p}_2 \hat{p}_4 \big) .\label{quantum_x_b}
\end{gather}
Here $\hat{p}_i$ is a center
\begin{gather*}
 \hat{p}_{ab} \hat{p}_i = \hat{p}_i \hat{p}_{ab}, \qquad \hat{p}_i \hat{p}_j = \hat{p}_j \hat{p}_i .
\end{gather*}

As a quantization of Theorem~\ref{thm:map_sphere}, we have the following.
\begin{Theorem} \label{thm:qW_for_sphere} We have an embedding
 \begin{gather*}
 \iota\colon \ \mathcal{X}^q(\Sigma_{0,4}) \to \mathcal{A}^q(Q_{\text{\rm oct}})
 \end{gather*}
 defined by
 \begin{alignat}{3}
& \iota(\hat{p}_1) = q^{-\frac{1}{4}} \sum_{\epsilon\in \{+,-\} } {Y_1}^{\frac{\epsilon}{2}} {Y_3}^{\frac{\epsilon}{2}} {Y_6}^{\frac{\epsilon}{2}} ,\qquad
 & &\iota(\hat{p}_2) = q^{-\frac{1}{4}} \sum_{\epsilon\in \{+,-\} } {Y_2}^{\frac{\epsilon}{2}} {Y_3}^{\frac{\epsilon}{2}} {Y_5}^{\frac{\epsilon}{2}} , & \nonumber\\
 & \iota(\hat{p}_3) = q^{-\frac{1}{4}} \sum_{\epsilon\in \{+,-\} } {Y_1}^{\frac{\epsilon}{2}} {Y_4}^{\frac{\epsilon}{2}} {Y_5}^{\frac{\epsilon}{2}} ,\qquad & &
 \iota(\hat{p}_4) = q^{-\frac{1}{4}} \sum_{\epsilon\in \{+,-\} } {Y_2}^{\frac{\epsilon}{2}} {Y_4}^{\frac{\epsilon}{2}} {Y_6}^{\frac{\epsilon}{2}} ,&\label{qp_YYY}
 \end{alignat}
 and
\begin{gather*}
\iota(\hat{p}_{12}) = \sum_{ ( \epsilon_5, \epsilon_6,\epsilon_1, \epsilon_2) \in \mathfrak{E} } q^{-\frac{1}{4} (\epsilon_5+\epsilon_6) (\epsilon_1+\epsilon_2) } {Y_5}^{\frac{\epsilon_5}{2}}
 {Y_6}^{\frac{\epsilon_6}{2}} {Y_1}^{\frac{\epsilon_1}{2}} {Y_2}^{\frac{\epsilon_2}{2}} , \\
 \iota( \hat{p}_{31} ) = \sum_{ (\epsilon_3, \epsilon_4, \epsilon_5, \epsilon_6) \in \mathfrak{E} } q^{-\frac{1}{4} (\epsilon_5+\epsilon_6) (\epsilon_3+\epsilon_4) }
 {Y_3}^{\frac{\epsilon_3}{2}} {Y_4}^{\frac{\epsilon_4}{2}} {Y_5}^{\frac{\epsilon_5}{2}} {Y_6}^{\frac{\epsilon_6}{2}} , \\
 \iota(\hat{p}_{23}) = \sum_{ (\epsilon_1, \epsilon_2, \epsilon_3, \epsilon_4) \in \mathfrak{E} } q^{-\frac{1}{4} (\epsilon_3+\epsilon_4) (\epsilon_1+\epsilon_2) } {Y_1}^{\frac{\epsilon_1}{2}}
 {Y_2}^{\frac{\epsilon_2}{2}} {Y_3}^{\frac{\epsilon_3}{2}} {Y_4}^{\frac{\epsilon_4}{2}} .
\end{gather*}
Here
\begin{gather*}
 \mathfrak{E} = \{ ( \epsilon, \eta, +, + ) \,|\, \epsilon , \eta \in \{+, - \} \} \cup \{ ( -, -, \epsilon, \eta ) \,|\, \epsilon , \eta \in \{+, - \}\} .
\end{gather*}
\end{Theorem}

It is noted that $Y^{\frac{1}{2}}$ for $q$-commuting variable $Y$ is such that $(Y^{\frac{1}{2}})^2= Y$, and that $(Y Y^\prime)^{\frac{1}{2}}$ can be consistently given. Hence the definitions in Theorem~\ref{thm:qW_for_sphere} can also be written, for examples, as
\begin{gather} \label{q_p1}
 \iota(\hat{p}_1) = q^{-\frac{1}{2}} ( Y_1 Y_3 Y_6 )^{\frac{1}{2}} + q^{\frac{1}{2}} ( Y_1 Y_3 Y_6 )^{-\frac{1}{2}} ,
\end{gather}
and
\begin{gather}
 \iota\big(\hat{p}_{12}\big) = q^{-1} ( Y_5 Y_6 )^{-\frac{1}{2}} ( Y_1 Y_2 )^{-\frac{1}{2}} + ( Y_5 Y_6 )^{-\frac{1}{2}} \big({Y_1}^{-1} Y_2 \big)^{\frac{1}{2}} + ( Y_5 Y_6)^{-\frac{1}{2}} \big( Y_1{Y_2}^{-1}\big)^{\frac{1}{2}} \nonumber\\
\hphantom{\iota\big(\hat{p}_{12}\big) =}{} + q (Y_5 Y_6)^{-\frac{1}{2}} ( Y_1 Y_2)^{\frac{1}{2}} + \big( {Y_5}^{- 1} Y_6 \big)^{\frac{1}{2}} (Y_1 Y_2)^{\frac{1}{2}}\nonumber \\
\hphantom{\iota\big(\hat{p}_{12}\big) =}{} + \big( Y_5 {Y_6}^{-1} \big)^{\frac{1}{2}} ( Y_1 Y_2)^{\frac{1}{2}} + q^{-1} (Y_5 Y_6)^{\frac{1}{2}} (Y_1 Y_2 )^{\frac{1}{2}}\nonumber \\
\hphantom{\iota\big(\hat{p}_{12}\big)}{} = ( Y_5 Y_6 )^{-\frac{1}{2}} \big[ q \big( 1+q^{-1} {Y_1}^{-1} \big) \big( 1+q^{-1} {Y_2}^{-1} \big) + Y_5 + Y_6 + q^{-1} Y_5 Y_6 \big] ( Y_1 Y_2)^{\frac{1}{2}} . \label{q_p12}
 \end{gather}
Other realizations of $\hat{p}_i$, $\hat{p}_{31}$, and $\hat{p}_{23}$ are given similarly. Using $q$-commuting relations for $Y$-variables~\eqref{quantum_Y}, we can check~\eqref{center_x_b} after tedious computations. See Appendix~\ref{sec:qW_proof} for a brief sketch.

Note that $\hat{p}_{ab}$ and $\hat{p}_i$ satisfying~\eqref{center_x_b} and~\eqref{quantum_x_b} are constructed by use of the double affine Hecke algebra of type $C^\vee C_1$ in~\cite{Oblom04a} for $q=1$ (see also~\cite{Koorn08a,Terwil13a}). In~\cite{BulloPrzyt99a}, the algebra was studied based on the Kauffman bracket skein module on the punctured sphere $\Sigma_{0,4}$.

\subsection*{Automorphism}
Quantizing the mutations~\eqref{RL_mutation} by replacing $\mu$ by $\mu^q$,
\begin{gather*}
\mathsf{R}= \sigma_{5,6} \sigma_{1,5} \sigma_{2,6} \mu^q_1 \mu^q_2 \qquad \text{and} \qquad \mathsf{L}=\sigma_{5,6} \sigma_{3,5} \sigma_{4,6} \mu^q_3 \mu^q_4,
\end{gather*} we have the $\mathrm{PSL}(2;\mathbb{Z})$ action on the $Y$-variables as
\begin{gather*}
 \mathsf{R}(\boldsymbol{Y}) =
 \begin{pmatrix}
 Y_5 \big( 1+q {Y_1}^{-1} \big)^{-1} \big( 1+q {Y_2}^{-1} \big)^{-1} \\
 Y_6 \big( 1+q {Y_1}^{-1} \big)^{-1} \big( 1+q {Y_2}^{-1} \big)^{-1} \\
 Y_3 ( 1+q Y_1 ) (1+q Y_2) \\
 Y_4 ( 1+q Y_1) ( 1+q Y_2) \\
 {Y_2}^{-1} \\
 {Y_1}^{-1}
 \end{pmatrix} , \\
 \mathsf{L}(\boldsymbol{Y}) =
 \begin{pmatrix}
 Y_1 \big( 1+q {Y_3}^{-1} \big)^{-1} \big( 1+q {Y_4}^{-1}\big)^{-1} \\
 Y_2 \big( 1+q {Y_3}^{-1} \big)^{-1} \big( 1+q {Y_4}^{-1} \big)^{-1} \\
 Y_5 ( 1+q Y_3)( 1+q Y_4) \\
 Y_6 ( 1+q Y_3)( 1+q Y_4) \\
 {Y_4}^{-1} \\
 {Y_3}^{-1}
 \end{pmatrix} .
\end{gather*}
These satisfy the braid relation, $\mathsf{R} \circ \mathsf{L}^{-1} \circ \mathsf{R} = \mathsf{L}^{-1} \circ \mathsf{R} \circ \mathsf{L}^{-1}$. Correspondingly we have the following modular group actions on $\mathcal{X}^q(\Sigma_{0,4})$.

\begin{Proposition} We have
 \begin{gather}
 \mathsf{R}\colon \
 \left( \begin{matrix}
 \hat{p}_1 \\
 \hat{p}_2 \\
 \hat{p}_3 \\
 \hat{p}_4
 \end{matrix} \right)
 \mapsto
 \left( \begin{matrix}
 \hat{p}_2 \\
 \hat{p}_1 \\
 \hat{p}_3 \\
 \hat{p}_4
 \end{matrix}
 \right), \qquad
 \left( \begin{matrix}
 \hat{p}_{12} \\
 \hat{p}_{31} \\
 \hat{p}_{23}
 \end{matrix} \right)
 \mapsto
 \left( \begin{matrix}
 \hat{p}_{12} \\
 \hat{p}_{23} \\
 q \hat{p}_{23} \hat{p}_{12} - q^2 \hat{p}_{31} - q \big( \hat{p}_1 \hat{p}_3 + \hat{p}_2 \hat{p}_4 \big)
 \end{matrix} \right) , \nonumber\\
 \mathsf{L}\colon \
 \left( \begin{matrix}
 \hat{p}_1 \\
 \hat{p}_2 \\
 \hat{p}_3 \\
 \hat{p}_4
 \end{matrix} \right)
 \mapsto
 \left( \begin{matrix}
 \hat{p}_3 \\
 \hat{p}_2 \\
 \hat{p}_1 \\
 \hat{p}_4
 \end{matrix} \right) ,
 \qquad
 \left( \begin{matrix}
 \hat{p}_{12} \\
 \hat{p}_{31} \\
 \hat{p}_{23}
 \end{matrix} \right)
 \mapsto
 \left( \begin{matrix}
 \hat{p}_{23} \\
 \hat{p}_{31} \\
 q \hat{p}_{31} \hat{p}_{23} - q^2 \hat{p}_{12} - q \big( \hat{p}_1 \hat{p}_2 + \hat{p}_3 \hat{p}_4 \big)
 \end{matrix} \right) . \label{RL_qp}
 \end{gather}
\end{Proposition}

See Appendix~\ref{sec:q_mutation} for computations.

\section{Quantization of degenerate character varieties}\label{sec:q_confluence}
We shall give degenerations of the quantized character variety $\mathcal{X}^q(\Sigma_{0,4})$. We define the quantized degenerate character variety $\mathcal{X}^q_J$ by algebra generated by $\hat{p}_{12}$, $\hat{p}_{31}$, $\hat{p}_{23}$ with a~quotient by
\begin{gather*} 
 W^q_J\big(\hat{p}_{12},\hat{p}_{31},\hat{p}_{23}; \hat{p}_1, \dots, \hat{p}_{|J|} \big) =0 ,
\end{gather*}
and $q$-commuting relations among $\hat{p}_{12}$, $\hat{p}_{31}$, $\hat{p}_{23}$ defined below. Here $\hat{p}_i$ is a center,
\begin{gather*} 
 \hat{p}_i \hat{p}_j = \hat{p}_j \hat{p}_i, \qquad \hat{p}_i \hat{p}_{12} = \hat{p}_{12} \hat{p}_{i} , \qquad \hat{p}_i \hat{p}_{31} = \hat{p}_{31} \hat{p}_{i} , \qquad \hat{p}_i \hat{p}_{23} = \hat{p}_{23} \hat{p}_{i} .
\end{gather*}
In the following, we explicitly give $\mathcal{X}^q_J$ as a reduction of $\mathcal{X}^q(\Sigma_{0,4})$ based on Fig.~\ref{fig:degenerate}. All are reductions of $\mathcal{X}^q(\Sigma_{0,4})$, and are quantizations of the degenerated character varieties $\mathcal{X}_J$ in the previous section. Our claim is the following as a quatinzation of Theorem~\ref{thm:map_sphere}.

\begin{Theorem} \label{thm:q_sphere} We have an embedding
 \begin{gather*}
 \iota\colon \ \mathcal{X}^q_J \to \mathcal{A}^q_J,
 \end{gather*}
 where $\mathcal{A}^q_J$ is the quantum cluster algebra associated to the quiver ${Q}_J$ in Fig.~{\rm \ref{fig:degenerate}}.
\end{Theorem}

\subsection[$\mathcal{P}_{(2,1,1)}$]{$\boldsymbol{\mathcal{P}_{(2,1,1)}}$}
We set $Y_6 = \kappa q ( Y_2 Y_4)^{-1}$, and take a limit $\kappa \to \infty$. We have
\begin{gather*}
 W^q_{(2,1,1)}\big(\hat{p}_{12}, \hat{p}_{31}, \hat{p}_{23} ; \hat{p}_1, \hat{p}_2, \hat{p}_3\big) = q^{-1} \hat{p}_{12} \hat{p}_{23} \hat{p}_{31} - \bigl[ q^{-2} {\hat{p}_{12}}^{2} + q^{-2} {\hat{p}_{31}}^{2} \\
 \qquad {} + q^{-1} \big( \hat{p}_1 \hat{p}_2 + \hat{p}_3 \big) \hat{p}_{12} + q^{-1} \big( \hat{p}_1 \hat{p}_3 + \hat{p}_2 \big) \hat{p}_{31} + q \hat{p}_1 \hat{p}_{23} + {\hat{p}_1}^{2} +
 \hat{p}_1 \hat{p}_2 \hat{p}_3 + 1 \bigr],
\end{gather*}
with
\begin{gather*}
 q \hat{p}_{12} \hat{p}_{31} -q^{-1} \hat{p}_{31} \hat{p}_{12} = \big( q - q^{-1} \big) \hat{p}_1 , \\
 q \hat{p}_{31} \hat{p}_{23} -q^{-1} \hat{p}_{23} \hat{p}_{31} = \big(q^2 - q^{-2}\big) \hat{p}_{12} + \big( q - q^{-1} \big) \big( \hat{p}_1 \hat{p}_2 + \hat{p}_3 \big), \\
 q \hat{p}_{23} \hat{p}_{12} -q^{-1} \hat{p}_{12} \hat{p}_{23} = \big( q^2 - q^{-2} \big) \hat{p}_{31} + \big( q - q^{-1} \big) \big( \hat{p}_1 \hat{p}_3 + \hat{p}_2 \big) .
\end{gather*}
We have a map
\begin{gather}
 \iota \big(\hat{p}_1\big) = \big( Y_1 {Y_2}^{-1} \big)^{\frac{1}{2}} \big( Y_3 {Y_4}^{-1} \big)^{\frac{1}{2}},\nonumber \\
 \iota\big(\hat{p}_2\big) = q^{-\frac{1}{4}} \sum_{\varepsilon=\pm 1} {Y_2}^{\frac{\varepsilon}{2}} {Y_3}^{\frac{\varepsilon}{2}} {Y_5}^{\frac{\varepsilon}{2}} ,\nonumber \\
\iota\big(\hat{p}_3\big) = q^{-\frac{1}{4}} \sum_{\varepsilon=\pm 1} {Y_1}^{\frac{\varepsilon}{2}} {Y_4}^{\frac{\varepsilon}{2}} {Y_5}^{\frac{\varepsilon}{2}} ,\nonumber \\
 \iota\big(\hat{p}_{12}\big) = q^{-\frac{1}{4}} {Y_1}^{\frac{1}{2}} {Y_4}^{-\frac{1}{2}} ( 1 + q Y_5 ) {Y_5}^{-\frac{1}{2}} ,\nonumber \\
 \iota\big(\hat{p}_{31}\big) = q^{-\frac{1}{4}} {Y_2}^{-\frac{1}{2}} {Y_3}^{-\frac{1}{2}} \big[ ( 1+Y_3) \big( 1+{Y_4}^{-1} \big) + q^{-1} {Y_4}^{-1} {Y_5}^{-1} \big] {Y_5}^{\frac{1}{2}} ,\nonumber \\
 \iota\big(\hat{p}_{23}\big) = ( Y_1 Y_2 )^{-\frac{1}{2}} \big[ q \big( 1 + q^{-1} {Y_3}^{-1} \big) \big( 1+q^{-1} {Y_4}^{-1} \big) + \big( Y_1 + Y_2 + q^{-1} Y_1 Y_2 \big) \big] ( Y_3 Y_4 )^{\frac{1}{2}} . \label{q_P211}
\end{gather}

\subsubsection*{Automorphism}
The automorphism is given from the quantization of the mutation~\eqref{G_211}, $\mathsf{G}^q=\sigma_{1,2} \sigma_{1,4} \sigma_{2,3}\mu^q_1 \mu^q_2$. We have an action on $Y$-variables as
\begin{gather*}
 \mathsf{G}^q \colon \
 \begin{pmatrix}
 Y_1 \\ Y_2 \\ Y_3 \\ Y_4 \\ Y_5
 \end{pmatrix}
 \mapsto
 \begin{pmatrix}
 Y_3 ( 1 + q Y_1 ) ( 1 + q Y_2 ) \\
 Y_4 ( 1 + q Y_1 ) ( 1 + q Y_2 ) \\
 {Y_2}^{-1} \\
 {Y_1}^{-1} \\
 Y_5 \big( 1 + q {Y_1}^{-1} \big)^{-1} \big( 1 + q {Y_2}^{-1} \big)^{-1}
 \end{pmatrix} .
\end{gather*}
The action on the degenerated character variety is read as
\begin{gather*}
 \mathsf{G}^q\colon \
 \begin{pmatrix}
 \hat{p}_1 \\ \hat{p}_2 \\ \hat{p}_3
 \end{pmatrix}
 \mapsto
 \begin{pmatrix}
 \hat{p}_1 \\ \hat{p}_3 \\ \hat{p}_2
 \end{pmatrix} , \qquad
 \begin{pmatrix}
 \hat{p}_{12} \\
 \hat{p}_{31} \\
 \hat{p}_{23}
 \end{pmatrix}
 \mapsto
 \begin{pmatrix}
 q^{-1} \hat{p}_{12} \hat{p}_{23} - q^{-2} \hat{p}_{31} - q^{-1} \big( \hat{p}_1 \hat{p}_3 + \hat{p}_2 \big) \\
 \hat{p}_{12} \\
 \hat{p}_{23}
 \end{pmatrix} .
\end{gather*}

\subsection[$\mathcal{P}_{(3,1)}$]{$\boldsymbol{\mathcal{P}_{(3,1)}}$}
We set $Y_1 = \kappa q (Y_4 Y_5)^{-1}$ in $\mathcal{P}_{(2,1,1)}$~\eqref{q_P211}, and take a limit $\kappa\to\infty$. We obtain
\begin{gather*}
 W^q_{(3,1)}\big(\hat{p}_{12}, \hat{p}_{31}, \hat{p}_{23} ; \hat{p}_1, \hat{p}_2\big) \\
\qquad{} = q^{-1} \hat{p}_{12} \hat{p}_{23} \hat{p}_{31} - \big[ q^{-2} {\hat{p}_{12}}^{2} + q^{-1} \big( \hat{p}_1 \hat{p}_2 + 1 \big) \hat{p}_{12} + q^{-1} \hat{p}_1 \hat{p}_{31} + q \hat{p}_1 \hat{p}_{23}
 + \hat{p}_1 \big( \hat{p}_1 + \hat{p}_2 \big) \big],
\end{gather*}
with
\begin{gather*}
q \hat{p}_{12} \hat{p}_{31} - q^{-1} \hat{p}_{31} \hat{p}_{12} = \big(q - q^{-1} \big) \hat{p}_1 , \\
q \hat{p}_{31} \hat{p}_{23} - q^{-1} \hat{p}_{23} \hat{p}_{31} = \big( q^2 - q^{-2} \big) \hat{p}_{12} + \big(q-q^{-1}\big) \big( \hat{p}_1 \hat{p}_2 + 1 \big) , \\
q \hat{p}_{23} \hat{p}_{12} - q^{-1} \hat{p}_{12} \hat{p}_{23} = \big( q-q^{-1} \big) \hat{p}_1 .
\end{gather*}
We have
\begin{gather}
\iota\big(\hat{p}_1\big) = q^{-\frac{1}{4}} {Y_2}^{-\frac{1}{2}} {Y_3}^{\frac{1}{2}} {Y_4}^{-1} {Y_5}^{-\frac{1}{2}} , \nonumber\\
 \iota\big(\hat{p}_2\big) = q^{-\frac{1}{4}} \big( {Y_2}^{\frac{1}{2}} {Y_3}^{\frac{1}{2}} {Y_5}^{\frac{1}{2}} + {Y_2}^{-\frac{1}{2}} {Y_3}^{-\frac{1}{2}} {Y_5}^{-\frac{1}{2}} \big), \nonumber\\
 \iota\big(\hat{p}_{12}\big) = {Y_4}^{-1} \big( 1 + q^{-1} {Y_5}^{-1} \big) ,\nonumber \\
 \iota\big(\hat{p}_{31}\big) = q^{-\frac{1}{4}} {Y_2}^{-\frac{1}{2}} {Y_3}^{-\frac{1}{2}} \big[ ( 1+Y_3)\big(1+ {Y_4}^{-1} \big) + q^{-1} {Y_4}^{-1} {Y_5}^{-1} \big] {Y_5}^{\frac{1}{2}} ,\nonumber \\
 \iota\big(\hat{p}_{23}\big) = q^{\frac{3}{4}} {Y_2}^{-\frac{1}{2}} \big( 1 + q^{-1} Y_2 \big) {Y_3}^{\frac{1}{2}} {Y_5}^{-\frac{1}{2}} . \label{q_P31}
\end{gather}

\subsubsection*{Automorphism}
A quantization of the mutation~\eqref{G_31}, $\mathsf{G}^q=\sigma_{2,5} \sigma_{3,4} \mu^q_3 \mu^q_4$, gives the action on $Y$-variables as
\begin{gather*}
 \mathsf{G}^q \colon \
 \begin{pmatrix}
 Y_2 \\ Y_3 \\ Y_4 \\ Y_5
 \end{pmatrix}
 \mapsto
 \begin{pmatrix}
 Y_5 ( 1 + q Y_3 ) ( 1 + q Y_4 ) \\
 {Y_4}^{-1} \\
 {Y_3}^{-1} \\
 Y_2 \big( 1 + q {Y_3}^{-1} \big)^{-1} \big( 1 + q {Y_4}^{-1} \big)^{-1}
 \end{pmatrix} ,
\end{gather*}
which induces an action on $\mathcal{X}^q_{(3,1)}$ as
\begin{gather*}
 \mathsf{G}^q\colon \
 \begin{pmatrix}
 \hat{p}_1 \\
 \hat{p}_2
 \end{pmatrix}
 \mapsto
 \begin{pmatrix}
 \hat{p}_1 \\
 \hat{p}_2
 \end{pmatrix} , \qquad
 \begin{pmatrix}
 \hat{p}_{12} \\
 \hat{p}_{31} \\
 \hat{p}_{23}
 \end{pmatrix}
 \mapsto
 \begin{pmatrix}
 q^{-1} \big( \hat{p}_{23} \hat{p}_{31} - q^{-1} \hat{p}_{12} - \hat{p}_1 \hat{p}_2 -1 \big) \\
 \hat{p}_{23} \\
 \hat{p}_{31}
 \end{pmatrix} .
\end{gather*}

\subsection[$\mathcal{P}_{(2,2)}$]{$\boldsymbol{\mathcal{P}_{(2,2)}}$}
We put $Y_5=\kappa q ( Y_1 Y_4 )^{-1}$ in $\mathcal{P}_{(2,1,1)}$~\eqref{q_P211}, and take a limit $\kappa\to\infty$. We have
\begin{gather*}
 W^q_{(2,2)} \big( \hat{p}_{12}, \hat{p}_{31}, \hat{p}_{23}; \hat{p}_1, \hat{p}_2\big) \\
 \qquad {}=q^{-1} \hat{p}_{12} \hat{p}_{23} \hat{p}_{31} - \big[ q^{-2} {\hat{p}_{12}}^{2} +q^{-2} {\hat{p}_{31}}^{2}
 + q^{-1} \big( \hat{p}_1 \hat{p}_2 + 1 \big) \hat{p}_{12} + q^{-1} \big( \hat{p}_1 + \hat{p}_2 \big) \hat{p}_{31} +\hat{p}_1 \hat{p}_2 \big],
\end{gather*}
with
\begin{gather*}
q \hat{p}_{12} \hat{p}_{31} - q^{-1} \hat{p}_{31} \hat{p}_{12} =0 , \\
q \hat{p}_{31} \hat{p}_{23} - q^{-1} \hat{p}_{23} \hat{p}_{31} = \big( q^2 - q^{-2} \big) \hat{p}_{12} + \big(q-q^{-1}\big) \big( \hat{p}_1 \hat{p}_2 + 1 \big) , \\
q \hat{p}_{23} \hat{p}_{12} - q^{-1} \hat{p}_{12} \hat{p}_{23} = \big( q^2 - q^{-2} \big) \hat{p}_{31} + \big(q-q^{-1}\big) \big( \hat{p}_1+ \hat{p}_2 \big) .
\end{gather*}
An embedding is given by
\begin{gather}
\iota\big(\hat{p}_1\big) = \big(Y_1 {Y_2}^{-1}\big)^{\frac{1}{2}} \big( Y_3 {Y_4}^{-1}\big)^{\frac{1}{2}} , \nonumber\\
\iota\big(\hat{p}_2\big) = \big({Y_1}^{-1} {Y_2}\big)^{\frac{1}{2}} \big( Y_3 {Y_4}^{-1}\big)^{\frac{1}{2}} , \nonumber\\
\iota\big(\hat{p}_{12}\big) = {Y_4}^{-1} , \nonumber\\
\iota\big(\hat{p}_{31}\big) = q^{-1} (Y_1 Y_2)^{-\frac{1}{2}} (Y_3 Y_4)^{-\frac{1}{2}} ( 1 + q Y_3 ) \big( 1 + q^{-1} {Y_4}^{-1} \big) , \nonumber\\
\iota\big(\hat{p}_{23}\big) = (Y_1 Y_2 )^{-\frac{1}{2}} \big[ q \big( 1 + q^{-1} {Y_3}^{-1} \big) \big( 1 + q^{-1} {Y_4}^{-1}\big) + \big( Y_1 + Y_2 + q^{-1} Y_1 Y_2 \big)\big] (Y_3 Y_4)^{\frac{1}{2}} . \label{q_P22}
\end{gather}

\subsubsection*{Automorphism}
A quantization of the mutation~\eqref{G_22}, $\mathsf{G}^q = \sigma_{1,3} \sigma_{2,4} \mu^q_1 \mu^q_2 $, acts on $Y$-variables as
\begin{gather*}
 \mathsf{G}^q\colon \
 \begin{pmatrix}
 Y_1 \\ Y_2 \\ Y_3 \\Y_4
 \end{pmatrix}
 \mapsto
 \begin{pmatrix}
 Y_3 ( 1+q Y_1 ) ( 1+q Y_2 ) \\
 Y_4 ( 1+q Y_1 ) ( 1+q Y_2 ) \\
 {Y_1}^{-1} \\
 {Y_2}^{-1}
 \end{pmatrix} .
\end{gather*}
This action induces the following action;
\begin{gather*}
 \mathsf{G}^q \colon \
 \begin{pmatrix}
 \hat{p}_1 \\
 \hat{p}_2
 \end{pmatrix}
 \mapsto
 \begin{pmatrix}
 \hat{p}_2 \\
 {\hat{p}_1}^{-1}
 \end{pmatrix} ,
 \qquad
 \begin{pmatrix}
 \hat{p}_{12} \\
 \hat{p}_{31} \\
 \hat{p}_{23}
 \end{pmatrix}
 \mapsto
 \begin{pmatrix}
 q^{-1} {\hat{p}_1}^{-1} \big( \hat{p}_{12} \hat{p}_{23} - q^{-1} \hat{p}_{31} - \hat{p}_1 - \hat{p}_2 \big) \\
 {\hat{p}_1}^{-1} \hat{p}_{12} \\
 \hat{p}_{23}
 \end{pmatrix} .
\end{gather*}

\subsection[$\mathcal{P}_{(2,1,1)}^{\text{deg}}$]{$\boldsymbol{\mathcal{P}_{(2,1,1)}^{\text{deg}}}$}
We renormalize $Y_2$ and $Y_3$ as $\frac{Y_2}{\varepsilon}$ and $\varepsilon Y_3$ respectively. We take $\varepsilon \to 0 $, and introduce a new variable by $Y_{2^\prime} = Y_2 Y_3$. We get
\begin{gather*}
 W^q_{(2,1,1)^{\text{deg}}}\big(\hat{p}_{12}, \hat{p}_{31}, \hat{p}_{23} ; \hat{p}_2, \hat{p}_3\big) \\
 \qquad{}= q^{-1} \hat{p}_{12} \hat{p}_{23} \hat{p}_{31} - \big[ q^{-2} {\hat{p}_{12}}^{2} + q^{-2} {\hat{p}_{31}}^{2} + q^{-1} \hat{p}_3 \hat{p}_{12} + q^{-1} \hat{p}_2 \hat{p}_{31} + 1 \big],
\end{gather*}
with
\begin{gather*}
q \hat{p}_{12} \hat{p}_{31} - q^{-1} \hat{p}_{31} \hat{p}_{12} =0 , \\
q \hat{p}_{31} \hat{p}_{23} - q^{-1} \hat{p}_{23} \hat{p}_{31} = \big(q^2 - q^{-2}\big) \hat{p}_{12} + \big(q-q^{-1}\big) \hat{p}_3 , \\
q \hat{p}_{23} \hat{p}_{12} - q^{-1} \hat{p}_{12} \hat{p}_{23} = \big(q^2 - q^{-2}\big) \hat{p}_{31} + \big(q-q^{-1}\big) \hat{p}_2 .
\end{gather*}
The map is given by
\begin{gather*}
\iota\big(\hat{p}_2\big) = (Y_{2^\prime} Y_5)^{\frac{1}{2}} + (Y_{2^\prime} Y_5)^{-\frac{1}{2}} , \\
\iota\big(\hat{p}_3\big) = q^{-\frac{1}{4}} \sum_{\varepsilon=\pm 1} {Y_1}^{\frac{\varepsilon}{2}} {Y_4}^{\frac{\varepsilon}{2}} {Y_5}^{\frac{\varepsilon}{2}} , \\
\iota\big(\hat{p}_{12}\big) = q^{-\frac{1}{4}} {Y_1}^{\frac{1}{2}} {Y_4}^{-\frac{1}{2}} ( 1 + q Y_5 ) {Y_5}^{-\frac{1}{2}} , \\
\iota\big(\hat{p}_{31}\big) = {Y_{2^\prime}}^{-\frac{1}{2}} \big( 1 + {Y_4}^{-1} + q^{-1} {Y_4}^{-1} {Y_5}^{-1}\big) {Y_5}^{\frac{1}{2}} , \\
\iota\big(\hat{p}_{23}\big) = q^{\frac{1}{4}} {Y_1}^{-\frac{1}{2}} \big[ {Y_{2^\prime}}^{-\frac{1}{2}} \big( 1 + q^{-1} {Y_4}^{-1} \big) + \big( 1 + q^{-1} Y_1\big) {Y_{2^\prime}}^{\frac{1}{2}}\big] {Y_4}^{\frac{1}{2}} .
\end{gather*}

\subsubsection*{Automorphism}
A quantized mutation follows from a quantization of~\eqref{G_211deg}, $\mathsf{G}^q=\sigma_{1,2^\prime} \sigma_{1,4} \mu^q_1$, as
\begin{gather*}
 \mathsf{G}^q\colon \
 \begin{pmatrix}
 Y_1 \\ Y_{2^\prime} \\ Y_4 \\ Y_5
 \end{pmatrix}
 \mapsto
 \begin{pmatrix}
 Y_{2^\prime} ( 1 + q Y_1) \\
 Y_4 ( 1 + q Y_1) \\
 {Y_1}^{-1} \\
 Y_5 \big( 1 + q {Y_1}^{-1} \big)^{-1}
 \end{pmatrix} ,
\end{gather*}
which induces
\begin{gather*}
 \mathsf{G}^q \colon \
 \begin{pmatrix}
 \hat{p}_2 \\
 \hat{p}_3
 \end{pmatrix}
 \mapsto
 \begin{pmatrix}
 \hat{p}_3 \\
 \hat{p}_2
 \end{pmatrix} , \qquad
 \begin{pmatrix}
 \hat{p}_{12} \\
 \hat{p}_{31} \\
 \hat{p}_{23}
 \end{pmatrix}
 \mapsto
 \begin{pmatrix}
 q \big( \hat{p}_{23} \hat{p}_{12} - q \hat{p}_{31} -\hat{p}_2 \big) \\
 \hat{p}_{12} \\
 \hat{p}_{23}
 \end{pmatrix} .
\end{gather*}

\subsection[$\mathcal{P}_{(4)A}$]{$\boldsymbol{\mathcal{P}_{(4)A}}$}
We put $Y_3 = \kappa q (Y_5 Y_2)^{-1}$ in $\mathcal{P}_{(3,1)}$~\eqref{q_P31}. Sending $\kappa\to\infty$, we get
\begin{gather*}
 W^q_{(4)A}\big( \hat{p}_{12}, \hat{p}_{31}, \hat{p}_{23}; \hat{p}_1 \big)
 = q^{-1} \hat{p}_{12} \hat{p}_{23} \hat{p}_{31} - \big[ q^{-1} \hat{p}_1 \hat{p}_{12} + q^{-1} \hat{p}_1 \hat{p}_{31} + q \hat{p}_1 \hat{p}_{23} + \hat{p}_1 \big( \hat{p}_1 + 1\big)\big] ,
\end{gather*}
with
\begin{gather*}
 q \hat{p}_{12} \hat{p}_{31} - q^{-1} \hat{p}_{31} \hat{p}_{12} = q \hat{p}_{31} \hat{p}_{23} - q^{-1} \hat{p}_{23} \hat{p}_{31} = q \hat{p}_{23} \hat{p}_{12} - q^{-1} \hat{p}_{12} \hat{p}_{23}
= \big( q - q^{-1}\big) \hat{p}_1 .
\end{gather*}
Here we have
\begin{gather}
 \iota(\hat{p}_1) = q^{-1} {Y_2}^{-1}{Y_4}^{-1}{Y_5}^{-1} ,\nonumber \\
\iota(\hat{p}_{12}) = {Y_4}^{-1} \big( 1 + q^{-1} {Y_5}^{-1}\big) , \nonumber\\
 \iota(\hat{p}_{31}) = {Y_2}^{-1}\big( 1 + q^{-1} {Y_4}^{-1}\big) , \nonumber\\
 \iota(\hat{p}_{23}) =\big( 1 + q {Y_2}^{-1}\big) {Y_5}^{-1} . \label{q_P4A}
\end{gather}

\subsection[$\mathcal{P}_{(4)B}$]{$\boldsymbol{\mathcal{P}_{(4)B}}$}
We set $Y_5 = \kappa q (Y_2 Y_3)^{-1}$ in $\mathcal{P}_{(3,1)}$~\eqref{q_P31}, and take a limit $\kappa\to\infty$. We get
\begin{gather*}
 W^q_{(4)B}\big( \hat{p}_{12}, \hat{p}_{31}, \hat{p}_{23}; \hat{p}_1 \big) = q^{-1} \hat{p}_{12} \hat{p}_{23} \hat{p}_{31} - \big[ q^{-2} {\hat{p}_{12}}^{2} + q^{-1}\big( \hat{p}_1 +1\big) \hat{p}_{12}
 + q^{-1} \hat{p}_1 \hat{p}_{31} + \hat{p}_1\big],
\end{gather*}
with
\begin{gather*}
 q \hat{p}_{12} \hat{p}_{31} - q^{-1} \hat{p}_{31} \hat{p}_{12} = 0, \\
 q \hat{p}_{31} \hat{p}_{23} - q^{-1} \hat{p}_{23} \hat{p}_{31} = \big(q^2 - q^{-2}\big) \hat{p}_{12} + \big( q- q^{-1}\big)\big( \hat{p}_1 + 1\big) , \\
 q \hat{p}_{23} \hat{p}_{12} - q^{-1} \hat{p}_{12} \hat{p}_{23} = \big( q - q^{-1}\big) \hat{p}_1 .
\end{gather*}
These are realized by
\begin{gather*}
 \iota\big(\hat{p}_1\big) = Y_3 {Y_4}^{-1} , \\
\iota\big(\hat{p}_{12}\big) = {Y_4}^{-1} , \\
 \iota\big(\hat{p}_{31}\big) = {Y_2}^{-1}\big( 1 + q^{-1} {Y_3}^{-1}\big)\big( 1 + q^{-1} {Y_4}^{-1}\big) , \\
\iota\big(\hat{p}_{23}\big) = \big( 1 + q {Y_2}^{-1}\big) {Y_3} .
\end{gather*}

\subsection[$\mathcal{P}_{(3,1)B}$]{$\boldsymbol{\mathcal{P}_{(3,1)B}}$}
We set $Y_4 = \kappa q( Y_5 Y_2)^{-1}$ in $\mathcal{P}_{(3,1)}$~\eqref{q_P31}, and take a limit $\kappa\to\infty$. We obtain
\begin{gather*}
 W^q_{(4)B}\big( \hat{p}_{12}, \hat{p}_{31}, \hat{p}_{23}; \hat{p}_1 \big) = q^{-1} \hat{p}_{12} \hat{p}_{23} \hat{p}_{31} -\big[ q^{-1} \hat{p}_{12} + q^{-1} \hat{p}_1 \hat{p}_{31} + q \hat{p}_1 \hat{p}_{23}
 + {\hat{p}_1}^{2} + 1\big],
\end{gather*}
with
\begin{gather*}
 q \hat{p}_{12} \hat{p}_{31} - q^{-1} \hat{p}_{31} \hat{p}_{12} =\big( q - q^{-1}\big) \hat{p}_1 , \\
 q \hat{p}_{31} \hat{p}_{23} - q^{-1} \hat{p}_{23} \hat{p}_{31} = q - q^{-1} , \\
 q \hat{p}_{23} \hat{p}_{12} - q^{-1} \hat{p}_{12} \hat{p}_{23} =\big( q - q^{-1}\big) \hat{p}_1 .
\end{gather*}
Here we have
\begin{gather*}
 \iota\big(\hat{p}_1\big)= q^{-\frac{1}{4}} {Y_2}^{\frac{1}{2}} {Y_3}^{\frac{1}{2}} {Y_5}^{\frac{1}{2}} , \\
 \iota\big(\hat{p}_{12}\big) = {Y_2} ( 1 + q Y_5 ) , \\
 \iota\big(\hat{p}_{31}\big) = q^{-\frac{1}{4}} {Y_2}^{-\frac{1}{2}} {Y_3}^{-\frac{1}{2}}( 1 + {Y_3}) {Y_5}^{\frac{1}{2}} , \\
 \iota\big(\hat{p}_{23}\big) = q^{\frac{3}{4}} {Y_2}^{-\frac{1}{2}} \big( 1 + q^{-1} {Y_2} \big) {Y_3}^{\frac{1}{2}} {Y_5}^{-\frac{1}{2}}.
\end{gather*}

\subsection[$\mathcal{P}_{(2,2)}^{\text{deg}}$]{$\boldsymbol{\mathcal{P}_{(2,2)}^{\text{deg}}}$}
We renormalize $Y_2$ and $Y_3$ by $\frac{Y_2}{\varepsilon}$ and $\varepsilon Y_3$ in~\eqref{q_P22} respectively. We take a limit $\varepsilon \to 0$, and introduce a new variable $Y_{2^\prime} = q^{-1} Y_2 Y_3$. We have
\begin{gather*}
 W^q_{(2,2)^{\text{deg}}}\big ( \hat{p}_{12}, \hat{p}_{31}, \hat{p}_{23}; \hat{p}_2 \big) = q^{-1} \hat{p}_{12} \hat{p}_{23} \hat{p}_{31} - \big[ q^{-2} {\hat{p}_{12}}^{2} +q^{-2} {\hat{p}_{31}}^{2} + q^{-1} \hat{p}_{12} + q^{-1} \hat{p}_2 \hat{p}_{31}\big] ,
\end{gather*}
with
\begin{gather*}
q \hat{p}_{12} \hat{p}_{31} - q^{-1} \hat{p}_{31} \hat{p}_{12} =0 , \\
q \hat{p}_{31} \hat{p}_{23} - q^{-1} \hat{p}_{23} \hat{p}_{31} = \big(q^2 - q^{-2}\big) \hat{p}_{12} + q-q^{-1} , \\
q \hat{p}_{23} \hat{p}_{12} - q^{-1} \hat{p}_{12} \hat{p}_{23} = \big(q^2 - q^{-2}\big) \hat{p}_{31} + \big(q-q^{-1}\big) \hat{p}_2 .
\end{gather*}
The map is given by
\begin{gather*}
\iota\big(\hat{p}_2\big) = q^{\frac{1}{4}} {Y_1}^{-\frac{1}{2}} {Y_{2^\prime}}^{\frac{1}{2}} {Y_4}^{-\frac{1}{2} } , \\
\iota\big(\hat{p}_{12}\big) = {Y_4}^{-1} , \\
\iota\big(\hat{p}_{31}\big) = q^{-\frac{3}{4}} {Y_1}^{-\frac{1}{2}} {Y_{2^\prime}}^{-\frac{1}{2}} {Y_4}^{-\frac{1}{2}}\big( 1 + q^{-1} {Y_4}^{-1}\big) , \\
\iota\big(\hat{p}_{23}\big) = q^{\frac{1}{4}} {Y_1}^{-\frac{1}{2}} \big[ {Y_{2^\prime}}^{-\frac{1}{2}} \big( 1 + q^{-1} {Y_4}^{-1}\big) + \big( 1+q^{-2} Y_1\big) {Y_{2^\prime}}^{\frac{1}{2}}\big] {Y_4}^{\frac{1}{2}} .
\end{gather*}

\subsection[$\mathcal{P}_{(2,2)}^{\text{deg}^2}$]{$\boldsymbol{\mathcal{P}_{(2,2)}^{\text{deg}^2}}$}
We renormalize $Y_1$ and $Y_{2^\prime}$ by $\frac{Y_1}{\varepsilon}$ and $\varepsilon Y_{2^\prime}$. We take $\varepsilon\to 0$, and set $Y_{2^{\prime \prime}}= q^{-1} Y_1 Y_{2^\prime}$. We obtain
\begin{gather*}
 W^q_{(2,2)^{\text{deg}^2}}\big( \hat{p}_{12}, \hat{p}_{31}, \hat{p}_{23} \big) = q^{-1} \hat{p}_{12} \hat{p}_{23} \hat{p}_{31} -\big[ q^{-2} {\hat{p}_{12}}^{2} + q^{-2} {\hat{p}_{31}}^{2} + q^{-1} \hat{p}_{12}\big] ,
\end{gather*}
with
\begin{gather*}
q \hat{p}_{12} \hat{p}_{31} - q^{-1} \hat{p}_{31} \hat{p}_{12} = 0 , \\
q \hat{p}_{31} \hat{p}_{23} - q^{-1} \hat{p}_{23} \hat{p}_{31} = \big(q^2 - q^{-2}\big) \hat{p}_{12} + q - q^{-1} , \\
q \hat{p}_{23} \hat{p}_{12} - q^{-1} \hat{p}_{12} \hat{p}_{23} = \big(q^2 - q^{-2}\big) \hat{p}_{31} .
\end{gather*}
The map is defined by
\begin{gather*}
\iota\big(\hat{p}_{12}\big) = {Y_4}^{-1} , \\
\iota\big(\hat{p}_{31}\big) = {Y_{2^{\prime\prime}}}^{-\frac{1}{2}} {Y_4}^{-\frac{1}{2}}\big( 1 + q^{-1} {Y_4}^{-1}\big) , \\
\iota\big(\hat{p}_{23}\big) = {Y_{2^{\prime\prime}}}^{-\frac{1}{2}} \big( 1 + q Y_4+ q^{-1} {Y_{2^{\prime\prime}}} Y_4\big) {Y_4}^{-\frac{1}{2}} .
\end{gather*}

\subsection[$\mathcal{P}_{(4)C}$]{$\boldsymbol{\mathcal{P}_{(4)C}}$}
We put $Y_4=\kappa q ( Y_5 Y_2)^{-1}$ in $\mathcal{P}_{(4)A}$~\eqref{q_P4A}, and take a limit $\kappa\to\infty$. We get
\begin{gather*}
 W_{(4)C}\big( \hat{p}_{12}, \hat{p}_{31}, \hat{p}_{23}\big) = q^{-1} \hat{p}_{12} \hat{p}_{23} \hat{p}_{31} -\big[ q^{-1} \hat{p}_{31} + q \hat{p}_{23} + 1\big],
\end{gather*}
with the $q$-commuting relations,
\begin{gather*}
q \hat{p}_{12} \hat{p}_{31} - q^{-1} \hat{p}_{31} \hat{p}_{12} = q-q^{-1} , \\
q \hat{p}_{31} \hat{p}_{23} - q^{-1} \hat{p}_{23} \hat{p}_{31} = 0 , \\
q \hat{p}_{23} \hat{p}_{12} - q^{-1} \hat{p}_{12} \hat{p}_{23} = q - q^{-1} .
\end{gather*}
The map is given by
\begin{gather*}
 \iota\big(\hat{p}_{12}\big) = {Y_2} ( 1 + q {Y_5} ) , \qquad \iota\big(\hat{p}_{31}\big) = {Y_2}^{-1} ,\qquad \iota\big(\hat{p}_{23}\big) = \big( 1 + q {Y_2}^{-1} \big) {Y_5}^{-1} .
\end{gather*}

\appendix
\section{Some computations}

\subsection{Quantized character variety}\label{sec:qW_proof}

We first note that, $\iota(\hat{p}_i)$ defined in~\eqref{qp_YYY} is a center of $\mathcal{A}^q(Q_{\text{oct}})$, due to that $Y_1 Y_3 Y_6$, $Y_2 Y_3 Y_5$, $Y_1 Y_4 Y_5$, $Y_2 Y_4 Y_6$ are central.

We check~\eqref{qW_is_zero} as follows. By use of the expressions~\eqref{q_p1} for $\hat{p}_i$, we have
\begin{gather*}
 \iota \big( \hat{p}_1 \hat{p}_2 + \hat{p}_3 \hat{p}_4 \big) =(Y_5 Y_6 )^{\frac{1}{2}} ( Y_1 Y_2 )^{-\frac{1}{2}} \bigl[ q^{-1} Y_1 Y_2 ( Y_3 + Y_4 ) \\
\hphantom{\iota \big( \hat{p}_1 \hat{p}_2 + \hat{p}_3 \hat{p}_4 \big) =}{} + q^2 \big( {Y_5}^{-1} + {Y_6}^{-1} \big) ( Y_1 + Y_2) + q^3 {Y_5}^{-1} {Y_6}^{-1} \big( {Y_3}^{-1} + {Y_4}^{-1} \big) \bigr] .
\end{gather*}
Using~\eqref{q_p12}, we then obtain
\begin{gather*}
 \iota \big( q^{-1} \hat{p}_{12} + \hat{p}_1 \hat{p}_2 + \hat{p}_3 \hat{p}_4 \big) \\
\qquad{} = ( Y_1 Y_2 )^{-\frac{1}{2}} \big\{ \big[\big( 1 + q^{-1} Y_1\big)\big( 1 + q^{-1} Y_2\big) + q^{-1}\big( {Y_3}^{-1} + {Y_4}^{-1}\big)\big] {Y_5}^{-1} {Y_6}^{-1} \\
 \qquad\quad{} + \big[ q^{-1} Y_1 Y_2 + Y_1 + Y_2\big] \big( {Y_5}^{-1} + {Y_6}^{-1}\big) + Y_1 Y_2 \big[ 1 + q^{-1} ( Y_3 + Y_4 )\big] \big\}( Y_5 Y_6)^{\frac{1}{2}} .
\end{gather*}
Further using the presentations~\eqref{q_p12} for $\hat{p}_{23}$ and $\hat{p}_{31}$, we get
\begin{gather}
\iota \big( \hat{p}_{12}\big\{ q^{-1} \hat{p}_{23} \hat{p}_{31} -q^{-2} \hat{p}_{12} - q^{-1}\big( \hat{p}_1 \hat{p}_2 + \hat{p}_3 \hat{p}_4\big)\big\}\big)\nonumber \\
 \qquad{} = \big[\big( 1 + q {Y_1}^{-1}\big)\big( 1 + q {Y_2}^{-1}\big) + q^{-1} ( Y_5 + Y_6 ) +q^{-2} Y_5 Y_6\big]\nonumber \\
 \qquad\quad{} \times \big[ q^{-6} {Y_3}^{-1} {Y_4}^{-1} {Y_5}^{-1} {Y_6}^{-1} + q^{-1} ( Y_1 +Y_2)( 1 + q Y_3)( 1 + q Y_4) + q^{-2} Y_1 Y_2 Y_3 Y_4\nonumber \\
 \qquad\quad{} + q \big( 1+ q^{-3}{Y_3}^{-1}\big)\big( 1 + q^{-3} {Y_4}^{-1}\big) \big\{ {Y_5}^{-1} + {Y_6}^{-1} + q ( 1 + q Y_3)( 1 + q Y_4)\big\}\big] . \label{qa_1}
 \end{gather}
Similar computations give the following
\begin{gather}
 \iota \big( \hat{p}_{31}\big\{ q^{-2} \hat{p}_{31} + q^{-1} \big( \hat{p}_1 \hat{p}_3 + \hat{p}_2 \hat{p}_4\big)\big\}\big)\nonumber \\
 \qquad{} = \big[ q^{-1} \big( 1 + q^{-1} Y_3\big)\big( 1 + q^{-1} Y_4\big) + {Y_5}^{-1} + {Y_6}^{-1} + q {Y_5}^{-1} {Y_6}^{-1}\big]\nonumber \\
 \qquad\quad{} \times \big[\big\{ q + q^{-2} \big( {Y_1}^{-1} + {Y_2}^{-1}\big)\big\} {Y_3}^{-1} {Y_4}^{-1} + \big\{ q\big( {Y_3}^{-1} + {Y_4}^{-1}\big) +q^2 {Y_3}^{-1} {Y_4}^{-1} \big\} ( Y_5 + Y_6 )\nonumber \\
\qquad\quad{} + \big\{ q^{-3}\big( 1 + q^3 {Y_3}^{-1}\big)\big( 1 + q^3 {Y_4}^{-1}\big) + q^2 ( Y_1 + Y_2 )\big\} Y_5 Y_6\big] , \label{qa_2}
\end{gather}
and
\begin{gather}
\iota\big( \hat{p}_{23}\big\{ q^2 \hat{p}_{23} + q\big( \hat{p}_1 \hat{p}_4 + \hat{p}_2 \hat{p}_3\big)\big\}\big) =
\big[ \big( 1 + q {Y_3}^{-1}\big)\big( 1 + q {Y_4}^{-1}\big) + q^{-1}( Y_1 + Y_2) + q^{-2} Y_1 Y_2\big]\nonumber \\
\qquad{} \times \big[ q^4 {Y_1}^{-1} {Y_2}^{-1} ( 1 + q Y_3 ) ( 1 + q Y_4 ) +\big( 1 + q^3\big( {Y_1}^{-1} + {Y_2}^{-1}\big)\big) Y_3 Y_4\nonumber \\
 \qquad{}+ q^2\big( {Y_1}^{-1} + {Y_2}^{-1}\big)( Y_3 + Y_4)+ q^{-1} Y_3 Y_4 ( Y_5 + Y_6 ) + q^3 {Y_1}^{-1} {Y_2}^{-1} \big( {Y_5}^{-1} + {Y_6}^{-1} \big)\big],\!\!\!\! \label{qa_3}
\end{gather}
Combining~\eqref{qa_1}--\eqref{qa_3}, we get
\begin{gather}
 \iota \big( \hat{p}_{12} \big\{ q^{-1} \hat{p}_{23} \hat{p}_{31} -q^{-2} \hat{p}_{12} - q^{-1} \big( \hat{p}_1 \hat{p}_2 + \hat{p}_3 \hat{p}_4\big) \big\}\nonumber \\
 \quad\quad{} - \hat{p}_{31} \big\{ q^{-2} \hat{p}_{31} + q^{-1} \big( \hat{p}_1 \hat{p}_3 + \hat{p}_2 \hat{p}_4 \big)\big\} - \hat{p}_{23} \big\{ q^2 \hat{p}_{23} + q \big( \hat{p}_1 \hat{p}_4 +
 \hat{p}_2 \hat{p}_3 \big) \big\} \big) \nonumber \\
 \quad{} = Y_1 {Y_2}^{-1} + {Y_1}^{-1} Y_2 + Y_3 {Y_4}^{-1} + {Y_3}^{-1} Y_4 + Y_5 {Y_6}^{-1} + {Y_5}^{-1} Y_6\nonumber \\
 \quad\quad {} + q^{-1} ( Y_1 + Y_2 ) ( Y_3 + Y_4)( Y_5 + Y_6) + q^{-1} \big( {Y_1}^{-1} + {Y_2}^{-1} \big) \big( {Y_3}^{-1} + {Y_4}^{-1}\big)\big( {Y_5}^{-1} + {Y_6}^{-1}\big)\nonumber \\
 \quad\quad {} + q^{-4} \big( {Y_1}^{-1} {Y_2}^{-1} {Y_3}^{-1} {Y_4}^{-1} {Y_5}^{-1} {Y_6}^{-1} + Y_1 Y_2 Y_3 Y_4 Y_5 Y_6\big) + 6 - q^2 - q^{-2} . \label{qa_4}
\end{gather}
On the other hand, we have from~\eqref{qp_YYY}
\begin{gather}
 \iota \big( \hat{p}_1 \hat{p}_2 \hat{p}_3 \hat{p}_4 \big) \label{qa_5}\\
 = \sum_{ \epsilon_1, \epsilon_2, \epsilon_3, \epsilon_4 \in \{+, -\} } q^{-1 - \frac{ \epsilon_1 \epsilon_2 + \epsilon_3 \epsilon_4 }{2} -\frac{( \epsilon_1 + \epsilon_2)
 \epsilon_3+\epsilon_4) }{2} } {Y_1}^{\frac{\epsilon_1+\epsilon_3}{2}} {Y_2}^{\frac{\epsilon_2+\epsilon_4}{2}} {Y_3}^{\frac{\epsilon_1+\epsilon_2}{2}} {Y_4}^{\frac{\epsilon_3+\epsilon_4}{2}}
 {Y_5}^{\frac{\epsilon_2+\epsilon_3}{2}} {Y_6}^{\frac{\epsilon_1+\epsilon_4}{2}} .\nonumber
\end{gather}
After some computations using~\eqref{qa_4} and~\eqref{qa_5}, we can indeed check the equality~\eqref{qW_is_zero}.

A set of the $q$-commuting relations~\eqref{quantum_x_b} can be checked similarly. By use of~\eqref{q_p12}, we have
\begin{gather*}
 \iota \big( q \hat{p}_{12} \hat{p}_{31} - q^{-1} \hat{p}_{31} \hat{p}_{12}\big) \\
\qquad{} = q ( Y_1 Y_2 )^{\frac{1}{2}} ( Y_3 Y_4)^{-\frac{1}{2}} \big[\big\{ q^{-3}\big( 1 + q^3 {Y_1}^{-1}\big)\big( 1 + q^3 {Y_2}^{-1}\big) + Y_5 + Y_6 + q^3 Y_5 Y_6\big\} \\
 \qquad\quad {} \times \big\{ q ( 1 + q Y_3)( 1 + q Y_4) + {Y_5}^{-1} +{Y_6}^{-1} + q^{-1} {Y_5}^{-1} {Y_6}^{-1}\big\} \\
\qquad\quad {} - \big\{ q^3 \big( 1 + q^{-3} Y_3 \big)\big( 1+ q^{-3} Y_4\big) + {Y_5}^{-1} + {Y_6}^{-1}+ q^{-3} {Y_5}^{-1} {Y_6}^{-1} \big\} \\
\qquad\quad {} \times \big\{ q^{-1} \big( 1 +q^{-1} {Y_1}^{-1} \big) \big( 1 + q^{-1} {Y_2}^{-1} \big) + Y_5 + Y_6 + q Y_5 Y_6 \big\}\big] \\
 \qquad{} = \big(q^2 - q^{-2}\big)( Y_1 Y_2)^{\frac{1}{2}} ( Y_3 Y_4t)^{- \frac{1}{2}} \\
\qquad\quad {} \times \big\{ q^{-1} Y_3 Y_4 + q^2 \big( {Y_1}^{-1} + {Y_2}^{-1}\big) Y_3 Y_4 + q^3 {Y_1}^{-1} {Y_2}^{-1} ( 1+ q Y_3 ) ( 1 + q Y_4)\big\} \\
\qquad\quad {} + \big( q- q^{-1}\big) \big\{ \big( {Y_1}^{\frac{1}{2}} {Y_2}^{-\frac{1}{2}} + {Y_1}^{-\frac{1}{2}} {Y_2}^{\frac{1}{2}}\big) \big(
 {Y_3}^{\frac{1}{2}} {Y_4}^{-\frac{1}{2}} + {Y_3}^{-\frac{1}{2}} {Y_4}^{\frac{1}{2}} \big) \\
\qquad\quad {} +q^{-2} ( Y_1 Y_2 Y_3 Y_4 )^{\frac{1}{2}} ( Y_5 + Y_6 ) + q^2 ( Y_1 Y_2 Y_3 Y_4)^{-\frac{1}{2}} \big( {Y_5}^{-1} + {Y_6}^{-1}\big) \big\} ,
\end{gather*}
which proves the first identity in~\eqref{quantum_x_b}. Others follow from cyclic permutation.

\subsection{Quantum mutations}\label{sec:q_mutation}
We shall check the $\mathsf{R}$-action in~\eqref{RL_qp}. We have $ Y_1 Y_3 Y_6 \xmapsto{\mathsf{R}} Y_2 Y_3 Y_5$, which proves $\hat{p}_1 \xmapsto{\mathsf{R}} \hat{p}_2$ from~\eqref{q_p1}. Actions on other $\hat{p}_i$ are given similarly. For $\hat{p}_{12}$, we use
\begin{gather*}
 Y_1 Y_2 \xmapsto{\mathsf{R}} Y_5 \big( 1 + q {Y_1}^{-1} \big)^{-1} \big( 1 + q {Y_2}^{-1}\big)^{-1} Y_6 \big( 1 + q {Y_1}^{-1} \big)^{-1} \big( 1 + q {Y_2}^{-1}\big)^{-1} \\
\qquad{} = \big\{ {Y_5}^{\frac{1}{2}} {Y_6}^{\frac{1}{2}} \big( 1 + q {Y_1}^{-1} \big)^{-1} \big( 1 + q {Y_2}^{-1} \big)^{-1} \big\}^2 .
\end{gather*}
Then we find
\begin{gather*}
( Y_5 Y_6)^{-\frac{1}{2}} \big[ {Y_1}^{-1} + {Y_2}^{-1} + q^{-1} {Y_1}^{-1} {Y_2}^{-1} + q \big( 1 + q^{-1} Y_5\big)\big( 1 + q^{-1} Y_6\big)\big] ( Y_1 Y_2)^{\frac{1}{2}} \\
\qquad{} \xmapsto{\mathsf{R}} ( Y_1 Y_2)^{\frac{1}{2}} \big[ {Y_5}^{-1} + {Y_6}^{-1} + q^{-1} \big( 1+ q {Y_1}^{-1} \big) \big( 1 + q^{-1} {Y_2}^{-1} \big) {Y_5}^{-1} {Y_6}^{-1} + q \big]
( Y_5 Y_6)^{-\frac{1}{2}} ,
\end{gather*}
which coincides with $\hat{p}_{12}$. One can also check $ \hat{p}_{31} \xmapsto{\mathsf{R}} \hat{p}_{23}$ in the same manner. For the action on~$\hat{p}_{23}$, we have
\begin{gather*}
(Y_1 Y_2)^{-\frac{1}{2}}\big( q^{-1} {Y_3}^{-1} {Y_4}^{-1} + {Y_3}^{-1} + {Y_4}^{-1} + q + Y_1 + Y_2 + q^{-1} Y_1 Y_2 \big) ( Y_3 Y_4)^{\frac{1}{2}} \\
\qquad{} \xmapsto{\mathsf{R}}( Y_5 Y_6)^{-\frac{1}{2}} \big[ q^{-1} \big\{ q \big(1+ q^{-1} {Y_1}^{-1}\big)\big( 1 + q^{-1} {Y_2}^{-1}\big) + Y_5 + Y_6 + q^{-1} Y_5 Y_6\big\} \\
 \qquad\quad{} \times \big\{ q \big( 1 +q^{-1} Y_1\big) \big( 1 + q^{-1} Y_2\big) + {Y_3}^{-1} +{Y_4}^{-1} + q^{-1} {Y_3}^{-1} {Y_4}^{-1} \big\} \\
\qquad\quad {} - q^{-2} \big\{ q {Y_5}^{-1} {Y_6}^{-1} + {Y_5}^{-1} + {Y_6}^{-1}+ q^{-1} ( 1 + q Y_3 ) ( 1 + q Y_4 ) \big\} \\
\qquad\quad {} -q^{-1} \big\{ q^{-1} \big( {Y_1}^{-1} + {Y_2}^{-1} \big) + \big( {Y_5}^{-1} + {Y_6}^{-1}\big) ( Y_3 + Y_4 ) + q^{-1} ( Y_1 + Y_2 ) Y_3 Y_4 \big\}\big]( Y_3 Y_4)^{\frac{1}{2}} .
 \end{gather*}
We see that this coincides with $q^{-1} \hat{p}_{12} \hat{p}_{23}-q^{-2} \hat{p}_{31}- q^{-1} ( \hat{p}_1 \hat{p}_3 + \hat{p}_2 \hat{p}_4 ) $, and that $q$-commutation relation~\eqref{quantum_x_b} gives the result. One can also check that the action is consistent with a quantized affine surface $W^q=0$. The action of $\mathsf{L}$ can be checked in the same manner.

\subsection*{Acknowledgments}
The author would like to thank Thang Le for communications during Workshop ``Low-Dimen\-sional Topology and Number Theory'' at Mathematisches Forschungsinstitut Oberwolfach in August 2017. He thanks the organizers for invitation. Thanks are also to the speakers of ``Geo\-metry of Moduli Spaces and Integrable Systems'' at Gakushuin University, Tokyo, in September 2017. This work is supported in part by JSPS KAKENHI Grant Number JP16H03927,
JP17K05239, JP17K18781, JP16H02143.

\pdfbookmark[1]{References}{ref}
\LastPageEnding

\end{document}